\documentclass[a4paper,11pt]{article}

\usepackage{subcaption} % To allow subfigures
\usepackage{jcappub} % For JCAP, contains already hyperref, color, graphicx and many others
\usepackage{aas_macros} % To read NASA/ADS bibtex
\usepackage{physics} % A lot of useful macros to write equations in physics
\usepackage{cleveref} % Clever references, eg \cref{eq:eq_1} -> eq. (1), \cref{sec:sec_1} -> sec. 1.

\usepackage{xspace}

\newcommand{\Refa}[1]{Ref.~{\cite{#1}}}
\newcommand{\Refs}[1]{Refs.~{\cite{#1}}}
\newcommand{\ie}{i.e.\ }
\newcommand{\eg}{e.g.\ }
\newcommand{\etc}{etc.\ }
\newcommand{\Mp}{M_{\scriptscriptstyle{\mathrm{Pl}}}}

\newcommand{\efolds}{$e$-folds\xspace}
\newcommand{\Pfpt}[1]{P_{\scriptscriptstyle{\mathrm{FPT}} , #1}}
\newcommand{\PfptV}[1]{P^{\mathrm{V}}_{\scriptscriptstyle{\mathrm{FPT}} , #1}}

\subheader{}
\title{Harvesting primordial black holes from stochastic trees with \texttt{FOREST}}

\author[a,b]{Chiara Animali,}
\author[a]{Pierre Auclair,}
\author[a,b]{Baptiste Blachier,}
\author[b]{Vincent Vennin}

\affiliation[a]{Cosmology, Universe and Relativity at Louvain (CURL), Institute of Mathematics and
	Physics, University of Louvain, 2 Chemin du Cyclotron, 1348 Louvain-la-Neuve, Belgium}

\affiliation[b]{Laboratoire de Physique de l'\'Ecole Normale Sup\'erieure, ENS, CNRS, Universit\'e PSL, Sorbonne Universit\'e, Universit\'e Paris Cit\'e, F-75005 Paris, France}

\emailAdd{chiara.animali@uclouvain.be}
\emailAdd{pierre.auclair@uclouvain.be}
\emailAdd{baptiste.blachier@uclouvain.be}
\emailAdd{vincent.vennin@ens.fr}

\date{today}

\begin{document}
\sloppy

\abstract{We introduce a novel framework to implement stochastic inflation on stochastic trees, modelling the inflationary expansion as a branching process. Combined with the $\delta N$ formalism, this allows us to generate real-space maps of the curvature perturbation that fully capture quantum diffusion and its non-perturbative backreaction during inflation. Unlike lattice methods, trees do not proceed on a fixed background since new spacetime units emerge dynamically as trees unfold, naturally incorporating metric fluctuations. The recursive structure of stochastic trees also offers remarkable numerical efficiency, and we develop the FOrtran Recursive Exploration of Stochastic Trees (\texttt{FOREST}) tool and demonstrate its performance. 
We show how primordial black holes blossom at unbalanced nodes of the trees, and how their mass distribution can be obtained while automatically accounting for the ``cloud-in-cloud'' effect. In the ``quantum-well'' toy model, we find broad mass distributions, with mild power laws terminated by exponential tails. We finally compare our results with existing approximations in the literature and discuss several prospects.}

\maketitle

%%%%%%%%%%%%%%%%%%%%%%%%%%%%%%%%%%%%%%%%%%%%%%%%%%%%%
%%%%%%%%%%%%%%%%%%%%%%%%%%%%%%%%%%%%%%%%%%%%%%%%%%%%%

\section{Introduction}
\label{sec:intro}

Observations of the universe at large cosmological distances reveal that density fluctuations have a small amplitude at large scales, 
and that they feature Gaussian, phase-coherent and scale-invariant statistics~\cite{SDSS:2005xqv, Planck:2018nkj, Ivanov:2019pdj}.
This is consistent with scenarios where cosmological perturbations arise from vacuum quantum fluctuations, 
stretched to astrophysical distances and amplified by gravitational instability~\cite{Mukhanov:1981xt, Mukhanov:1982nu, Starobinsky:1982ee, Guth:1982ec, Hawking:1982cz, Bardeen:1983qw} 
during an early era of accelerated expansion called inflation~\cite{Starobinsky:1980te, Sato:1981qmu, Guth:1980zm, Linde:1981mu, Albrecht:1982wi, Linde:1983gd}.

However, there have been recent hints for the existence of large (though rare) density fluctuations at small scales.
These include the existence of very massive galaxy clusters~\cite{Asencio:2020mqh}, 
the presence of galaxies and quasars at extremely high redshifts~\cite{2012ApJ...753..163G, Finkelstein:2013lfa, Eilers:2022mhw, 2023MNRAS.518.6011D}, 
and the observation in JWST~\cite{Harikane:2022rqt, 2022ApJ...940L..14N, 2022ApJ...940L..55F, 2023MNRAS.519.1201A, 2022ApJ...938L..13R, 2022ApJ...938L..15C, 2023ApJ...942L..27S, 2023MNRAS.518.4755A} of galaxies at very high redshift with large star-formation rate, 
advanced stellar maturity, or even harbouring black holes at a stage where they are not yet expected to do so if primordial perturbations are Gaussian and quasi scale invariant across all scales.
Large primordial density fluctuations may also lead to the formation of primordial black holes (PBHs), 
which could account for a fraction or all the dark matter, 
act as seeds for early structure formation, 
explain the existence of supermassive black holes in galactic nuclei, 
and constitute low-spin progenitors for the black-hole mergers detected by gravitational-wave experiments~\cite{Escriva:2022duf, Carr:2023tpt, LISACosmologyWorkingGroup:2023njw}.
Finally, large scalar perturbations can induce the production of gravitational waves through non-linear effects~\cite{Ananda:2006af, Baumann:2007zm}, 
which could explain the recent pulsar-timing array detections~\cite{EPTA:2023xxk, NANOGrav:2023hvm}.

If inflation is driven by a scalar field (the inflaton) rolling down a smooth potential function, in the attractor regime known as ``slow roll'' 
and if its dynamics is dominated by classical drift, then it is expected to give rise to small, quasi-Gaussian curvature perturbations.
However, features in the inflationary potential may make the inflaton deviate from the slow-roll attractor towards the end of inflation, 
or subject it to significant quantum diffusion, leading to enhanced density fluctuations at small scales.
In these cases, since large fluctuations are produced, standard cosmological perturbation theory often breaks down in the super-Hubble regime and non-perturbative techniques are instead required.
One such approach is the separate-universe picture~\cite{Salopek:1990jq, Sasaki:1995aw, Wands:2000dp, Lyth:2003im, Rigopoulos:2003ak, Lyth:2005fi, Artigas:2021zdk, Jackson:2023obv}, 
in which the causal structure of inflating spacetimes is used to describe it as an ensemble of independent Hubble-sized patches, locally homogeneous and isotropic, 
hence evolving according to local background equations of motion but with different initial conditions.
In this framework, the curvature perturbation $\zeta$ is identified with the local amount of expansion
\begin{equation}
	\label{eq:deltaN}
	\zeta(t,\vec{x})=N(t,\vec{x})-\overline{N}(t) \equiv \delta N\,,
\end{equation}
where $N(t,\vec{x})=\ln[a(t,\vec{x})]$ is the number of \efolds, $a(t,\vec{x})$ being the local value of the scale factor, and  $\overline{N}(t)$ is the unperturbed number of \efolds.
This is the $\delta N$ formalism~\cite{Starobinsky:1982ee, Starobinsky:1985ibc, Sasaki:1995aw, Sasaki:1998ug, Lyth:2004gb, Lyth:2005fi}.
In practice, the local amount of expansion, and its statistical properties, can be computed in the stochastic-inflation formalism~\cite{Starobinsky:1986fx}, 
where quantum fluctuations at small scales source the local background evolution as they cross out the Hubble radius during inflation, in the effective form of a random noise.
This makes the dynamics of background fields stochastic, and gives rise to the stochastic-$\delta N$ formalism~\cite{Enqvist:2008kt, Fujita:2013cna, Vennin:2015hra, Pattison:2017mbe}, 
where the curvature perturbation can be inferred from first-passage-time analysis.

This framework has been used to study the statistics of large fluctuations, which have been shown to feature exponential tails~\cite{Pattison:2017mbe, Ezquiaga:2019ftu, Panagopoulos:2019ail, Figueroa:2020jkf, Pattison:2021oen, Tomberg:2021xxv, Rigopoulos:2021nhv, Achucarro:2021pdh, Ezquiaga:2022qpw, Animali:2022otk, Cai:2022erk, Gow:2022jfb, Briaud:2023eae, Vennin:2024yzl, Inui:2024sce, Sharma:2024fbr}.
These are substantially heavier than Gaussian tails, which leads to more abundant large-curvature regions than what cosmological perturbation theory would predict.
However, determining the real-space profile of the curvature perturbation within the stochastic formalism remains challenging.
This difficulty arises because, when simulating realizations of the Langevin equation that governs the stochastic dynamics of the background fields, only a single worldline is tracked until inflation ceases.
Repeating this procedure generates an ensemble of such worldlines, but it provides no information about the spatial arrangement of the Hubble patches that terminate them on the end-of-inflation hypersurface.

In principle, in the separate-universe picture, the distance between two final Hubble patches is correlated with the time at which the worldlines they emerge from became independent.
Indeed, owing to the accelerated expansion, there is a time prior to which two such worldlines are distant by less than the Hubble radius, 
hence they belong to the same Hubble patch and follow the same Langevin realization.
When their distance grows larger than the Hubble radius they become independent, hence they follow separate realizations of the Langevin equation.
This is why patches that are closer one to the other on the end-of-inflation hypersurface tend to be more strongly correlated, 
since their worldlines became independent later, and they had less time to evolve away from their common past.
This can be used to compute correlation functions of the curvature perturbation, relying only on simulations of single realizations of the Langevin equation, 
first-passage-time analysis and conditional probabilities~\cite{Ando:2020fjm, Tada:2021zzj, Animali:2024jiz}.
Such methods have the advantage to provide analytical or semi-analytical results in some models, but they are so far limited to a subset of cosmological observables.
They also rely on approximations regarding the link between physical distances at the end of inflation and the field-space configuration when those scales emerged from a Hubble patch during inflation.
Indeed, the physical distance between the endpoints of two worldlines on the end-of-inflation hypersurface depends not only on the two trajectories, but also on the integrated expansion of all the patches in between them.
In classical setups, there is a one-to-one correspondence between this physical distance and the field-space configuration within the last common past patch of such worldlines, which does not hold in stochastic systems.
Different approximation schemes have been proposed, such as the ``backward-distribution approximation'' in \Refs{Ando:2020fjm, Tada:2021zzj} and the ``large-volume approximation'' in \Refa{Animali:2024jiz}, 
but the validity of these approximations remains to be further tested, and methods to go beyond are still missing.\\

In this article, we address these issues by proposing to implement the stochastic-$\delta N$ program on stochastic trees.
The idea is that, when a Hubble patch lying along one Langevin realization grows into two Hubble patches, it gives rise to two independent Langevin realizations.
This branching process is the elementary structure of a stochastic tree~\cite{Linde:1993xx,Jain:2019gsq}.
We show that the tree coordinate of a patch (that is, the direction its worldline followed at each branching process) can be mapped onto physical coordinates on the final hypersurface.
In application of the $\delta N$ formalism, the curvature perturbation on the leaves (namely, on the patches lying on the end-of-inflation hypersurface), is directly related to the time it took for them to reach the end of inflation.
Similarly, the time associated with branching nodes yields the curvature perturbation coarse-grained over its set of descendant leaves.
This makes the calculation of coarse-grained quantities straightforward, while automatically accounting for cloud-in-cloud effects in the context of structure (including PBH) formation.

Stochastic trees are commonly used in a number of contexts such as genetics and evolutionary biology where phylogenetic trees are often employed, 
in epidemiology where they are used to model the spread of infectious diseases, in computer science and machine learning where decision trees 
and random forests are used for classification and regression tasks, in finance and economics to model the pricing of options and other derivatives, 
in environmental science where trees are employed to simulate \eg the growth of forests and the spread of wildfires, 
in ecology to model the population dynamics of species within an ecosystem, in network theory, \etc \cite{Felsenstein2003,10.1093/oso/9780198545996.001.0001, breiman2001random, Glasserman, Hanski}.
This implies that a variety of mathematical and numerical tools have been developed to study and simulate stochastic trees.
By framing stochastic inflation as a tree problem, we can leverage these existing tools.

Note that other numerical approaches can be followed to simulate non-perturbative dynamics during inflation, 
such as numerical relativistic methods~\cite{East:2015ggf, Clough:2016ymm, Bloomfield:2019rbs, Joana:2020rxm, Corman:2022alv,Launay:2024qsm} 
(mostly employed to investigate the onset of inflation from non-homogeneous initial conditions) 
or lattice codes~\cite{Felder:2000hq, Frolov:2008hy, Easther:2010qz, Figueroa:2021yhd, Caravano:2021pgc}.

These approaches are numerically expensive, and do not include the effect of quantum diffusion, which is nonetheless essential in predicting the statistics of large fluctuations during inflation.
Recently however, lattice codes have been extended to include stochastic noise and perform the stochastic-$\delta N$ program. In particular, the code \texttt{STOLAS}~\cite{Mizuguchi:2024kbl} delivers three-dimensional maps of the curvature perturbation, where the effect of quantum diffusion and of non-linear evolution is included at large scales.
It thus represents a significant advance.

In lattice codes, simulations proceed on a fixed grid, \ie on flat hypersurfaces.
Metric perturbations can therefore not be included, and the stochastic lattice simulations only keep track of matter-field fluctuations.
The way the curvature perturbation is extracted in \Refa{Mizuguchi:2024kbl} is by evolving the grid until a time (well before the end of inflation) where the grid spacing exceeds the Hubble radius.
At that point, all nodes in the grid have become independent.
One can then solve individual stochastic realizations starting from each node, and record the number of \efolds until inflation ends.
Although this method is well adapted to some models, it relies on the assumption that inflation stops nowhere on the grid until nodes are evolved independently.
This either requires curvature fluctuations to be small, or to terminate the grid sufficiently before the end of inflation, hence to remove a sufficiently large range of small scales.
We will see that stochastic trees allow one to overcome these limitations: they are numerically much less expensive, given that the nodes never interact, and since they proceed on a non-rigid grid structure, the local (and coarse-grained) curvature perturbation can be straightforwardly computed.

This article is organized as follows.
In \cref{sec:tree}, we explain how stochastic trees can be used to describe the dynamics of inflating spacetimes, 
how they can be numerically implemented, and how the curvature perturbation can be extracted from them.
In \cref{sec:pbh}, we propose a criterion for PBH formation inspired by the well-known compaction function~\cite{Shibata:1999zs, Harada:2015yda, Musco:2018rwt} and expressed in terms of the quantities accessible during the production of a stochastic tree.
We test for PBH collapse at every node in the stochastic tree, automatically accounting for cloud-in-cloud.
As an illustration, we apply this formalism in \cref{sec:application} to reconstruct the mass fraction and mass distribution of PBHs in the quantum-well model of inflation, using populations of $10^{10}$ to $10^{11}$ trees.
In \cref{sec:discussion}, we consider how our results compare with previous analytical approximations and discuss the impact of discretization artefacts on our results.
We summarize our results and mention a few prospects in \cref{sec:conclusion}, and we end the paper with a few appendices to which technical details are deferred.

\section{Stochastic trees for inflation}
\label{sec:tree}

Although the stochastic-inflation formalism can be applied to multiple-field setups~\cite{Pinol:2020cdp}, 
and to models featuring deviations from the slow-roll attractor~\cite{Firouzjahi:2018vet, Pattison:2019hef, Ballesteros:2020sre, Raatikainen:2023bzk, Jackson:2024aoo}, 
for explicitness we will consider single-field slow-roll models, where the inflaton field $\phi$ coarse grained at the scale $R_\sigma=(\sigma H)^{-1}$ follows the Langevin and Friedmann equations
\begin{equation}
	\label{eq:Langevin}
	\dv{\phi}{N} = -\frac{V'(\phi)}{3H^2} + \frac{H}{2\pi}\xi(N)
	\quad\text{and}\quad H
    ^2=\frac{V(\phi)}{3\Mp^2}\,.
\end{equation}
In this expression, where the number of \efolds $N$ is used as a time variable~\cite{Finelli:2010sh}, $V(\phi)$ is the potential energy stored in the inflaton, and $\xi(N)$ is a normalized white Gaussian noise, such that $\ev{\xi(N)}=0$ and $\ev{\xi(N) \xi(N')} = \delta(N-N')$.
Along the slow-roll attractor the effective phase-space is one-dimensional~\cite{Grain:2017dqa}, but the extension of stochastic trees to higher-dimensional systems is straightforward.

\subsection{The tree structure of inflating spacetimes}
\label{sec:tree:structure}

\begin{figure}
	\centering
	\includegraphics[width=0.2\textwidth]{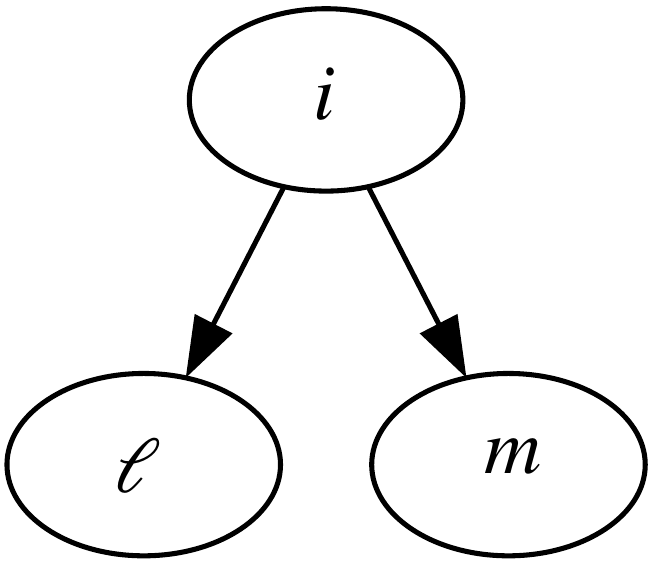}
	\caption{Elementary vertex of a stochastic tree, where one parent Hubble patch $i$ gives rise to two independent Hubble patches $\ell$ and $m$ after expanding for $\Delta N=\ln(2)/3$.}
	\label{fig:elementary:tree}
\end{figure}

In \cref{eq:Langevin}, $\phi$ represents the inflaton averaged over a physical region of size $R_\sigma=(\sigma H)^{-1}$ 
where $\sigma\ll 1$ is a fixed parameter that sets the ratio between the coarse-graining radius and the Hubble radius.
We call such a region a Hubble patch.
As the expansion proceeds, the physical volume of such a patch grows, and after a time
\begin{equation}
	\Delta N=\ln(2)/3
\end{equation}
it doubles in volume.
The situation is depicted in \cref{fig:elementary:tree} where a parent patch $i$ gives rise to two children patches $\ell$ and $m$.
In accelerating backgrounds, these two patches have no future causal contact, hence they evolve independently.
This gives rise to the ``separate-universe'' picture mentioned in \cref{sec:intro}.
This also implies that the field values in the patches $\ell$ and $m$, denoted $\phi_\ell$ and $\phi_m$ respectively, 
have to be evolved from the one in patch $i$, denoted $\phi_i$, using two distinct realizations of the Langevin equation~\eqref{eq:Langevin}.
In practice, the Langevin equation is thus solved twice over a duration $\Delta N$, from the same initial condition $\phi_i$, in order to get $\phi_\ell$ and $\phi_m$.

\begin{figure}
	\centering
	\includegraphics[width=0.5\textwidth]{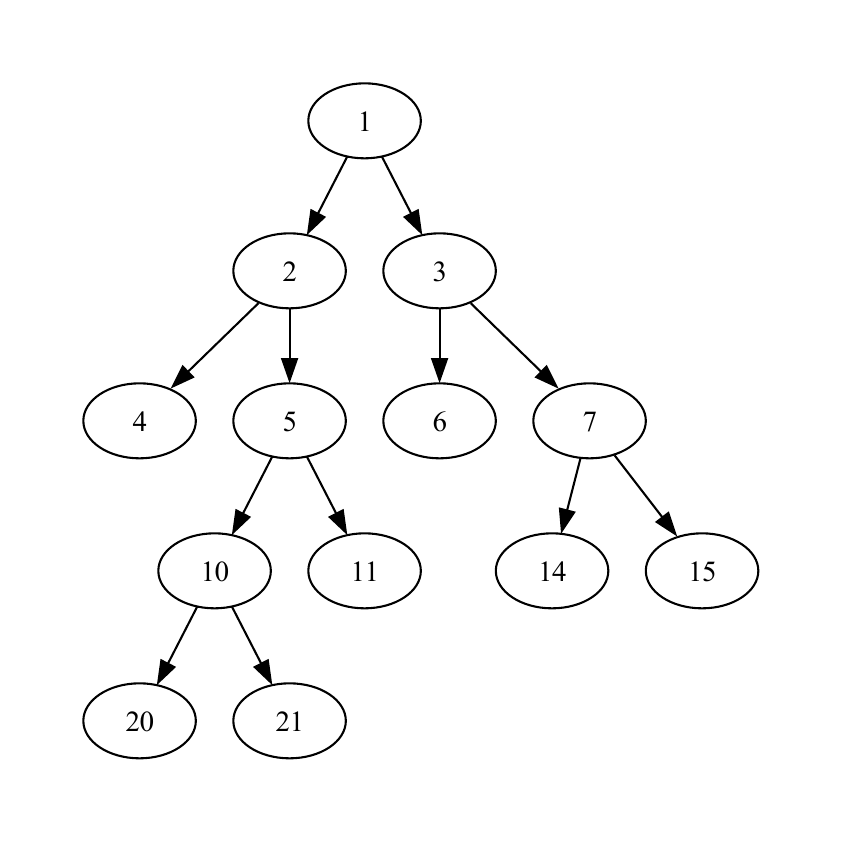}
	\caption{
        Example of a stochastic tree, made of multiple elementary vertices of the type shown in \cref{fig:elementary:tree}.
		The leaves are Hubble patches where inflation ends, and the patch labels correspond to their topological coordinates in the tree (see main text).
	}
	\label{fig:example:tree}
\end{figure}

When this procedure is iterated, the children patches give rise to two grand-children patches each, so on and so forth, leading to a binary tree structure.
An example of such a tree is displayed in \cref{fig:example:tree}.
It is made of ``nodes'' and ``branches'' that implement the causal structure of inflating spacetimes.

If inflation never stops, the tree grows indefinitely.
In practice however, inflation has to stop and the nodes where this occurs correspond to the ``leaves'' of the tree.
In the tree sketched in \cref{fig:example:tree}, the leaves are the nodes with labels $4$, $20$, $21$, $11$, $6$, $14$ and $15$.
In single-field slow-roll models, inflation most often stops by slow-roll violation, 
\ie when $\phi$ reaches a certain field value $\phi_\mathrm{end}$ where the potential function becomes too steep for inflation to proceed.
This is why, when evolving $\phi$ from a parent node $i$ to one of its child nodes $\ell$, if $\phi$ crosses $\phi_\mathrm{end}$ then the process is terminated and a leaf $L_\ell$ is created.
In units of the Hubble volume $V_\sigma=4\pi R_\sigma^3/3$ at the end of inflation, the physical volume of that leaf is given by (see \cref{footnote:split_then_grow})
\begin{equation}
	\label{eq:volume:leaf}
	V\left(L_\ell\right) = e^{3 \left(\mathcal{N}_{i\to \ell}-\Delta N\right)} \in [1/2, 1[ \, ,
\end{equation}
where $\mathcal{N}_{i\to \ell} < \Delta N$ denotes the number of \efolds realized along the final branch connecting the patches $i$ and $\ell$.

Stochastic trees of that type can be efficiently generated numerically using recursive methods, since each vertex gives rise to two subtrees that can be processed independently.
Once the routine implementing the elementary vertex depicted in \cref{fig:elementary:tree} is programmed, 
it shall call itself from each child node, until inflation ends at the tip of every branch and the entire tree is scanned.
We have developed \texttt{FOREST} (FOrtran Recursive Exploration of Stochastic Trees), a parallel Modern Fortran code using MPI+OpenMP that proceeds along these lines, 
and which is used to produce all numerical results presented in this paper.
\texttt{FOREST} takes as an input an inflationary potential function $V(\phi)$, its derivative $V'(\phi)$, and boundary conditions for the field configuration.
The Langevin equation~\eqref{eq:Langevin} is solved using the Euler–Maruyama method~\cite{Kloeden} with a varying step $\delta N$. 
Close to the end-of-inflation absorbing boundary, we enforce $\delta N < 3 [2\pi (\phi - \phi_\mathrm{end})] ^ 2 / V(\phi) / 5$ to limit the probability of barrier crossing to $5\sigma$ and avoid double crossings that spoil estimations of the first-passage time (see \cref{sec:convergence} for more details).
The value of the Hubble function is updated at each step $\delta N$ using Friedmann's equations.
We have successfully tested \texttt{FOREST} with populations of more than $10^{12}$ stochastic trees, shared across $2\times 256$ cores using MPI+OpenMP.
We report excellent scaling as a function of the number of cores as well as the number of trees (see \cref{sec:scaling}).
\texttt{FOREST} is publicly accessible under GPLv3 license \cite{auclair_2025_15235932}.

For practical purposes, each node of a given tree can be uniquely labelled using its ``trajectory'' in the binary tree.
Starting from the root node $1$, one may label the descendants of node $n$ as $2n$ for the leftward node and $2n+1$ for the rightward node.
As a consequence, the position of a node $n$ in the tree can be directly read from its representation in binary.
Consider for instance the patch labelled $11$ in \cref{fig:example:tree}, which reads as $1011_2$ in binary.
Starting from the parent patch labelled $1$, the path goes left ($0$), then right ($1$), then right ($1$) again.
This path may be represented as $(1, 0, 1, 1)$, which is indeed the representation of $11$ in binary.
In this way, a tree is entirely characterized by a set of node labels, and by the volume~\eqref{eq:volume:leaf} associated to each leaf.

\subsection{Curvature perturbation on the final hypersurface}
\label{sec:zeta}

Following the $\delta N$ formalism, the curvature perturbation corresponds to the local fluctuation in the amount of expansion, see \cref{sec:intro}.
In the tree structure introduced above, it can be computed as follows.

For each node $i$ in the tree, let $V_i$ be the physical volume emerging from that node, 
and let $W_i$ be the number of \efolds realized from that node, volume-averaged over all its child leaves.
In other words, if $\mathcal{L}_i$ denotes the set of leaves descending from node $i$, then one has
\begin{equation}\label{eq:Vi:Wi}
	V_i = \sum_{j \in \mathcal{L}_i} V(L_j) \quad \text{and} \quad W_i = \frac{1}{V_i} \sum_{j\in\mathcal{L}_i} V(L_j) \mathcal{N}_{i \to j}\, .
\end{equation}
Here, $V(L_j)$ is the volume of each leaf, computed according to \cref{eq:volume:leaf}, 
and $\mathcal{N}_{i\to j}$ is the number of \efolds realized on the path that connects the nodes $i$ and $j$.

Let us consider a late-time observer performing measurements in a bounded region of the final hypersurface.
That region contains a subset of leaves of the whole inflationary tree, 
and the lowest common ancestor of those leaves defines the primeval patch of the observer.
The leaves emerging from the primeval patch correspond to the ``observable universe'' of that observer, while the other leaves are inaccessible to them.
For explicitness, let us consider the case where the primeval patch is the root of the tree and is given the label $1$.
When measuring the curvature perturbation, the observer detects deviations of $\mathcal{N}$ from the mean value computed within its observable universe.
In each leaf $L_j$ they thus measure
\begin{equation}\label{eq:zeta:end:tree}
	\zeta_{V_j}(\vec{x}_j)=\mathcal{N}_{1\to j}-W_1\, .
\end{equation}
Here, $\vec{x}_j$ labels the position of the leaf $L_j$, 
and $\zeta_{V_j}$ corresponds to the curvature perturbation coarse-grained over a volume $V_j=V(L_j)$ 

The curvature perturbation can also be coarse-grained over volumes larger than individual leaves.
Consider for instance the set of leaves $\mathcal{L}_i$ emerging from a node $i$.
The volume-averaged value of $\zeta_{V_j}$ across that set corresponds to the curvature perturbation coarse-grained over the volume $V_i$.
In other words,
\begin{equation}
\begin{aligned}
    \label{eq:zeta:N:W}
    \zeta_i \equiv \zeta_{V_i}(\vec{x}_i)
    & = \frac{1}{V_i}\sum_{j \in \mathcal{L}_i} V_j \left(\mathcal{N}_{1\to j}-W_1\right)
    = \frac{1}{V_i}\sum_{j \in \mathcal{L}_i} V_j \left(\mathcal{N}_{1\to i}+\mathcal{N}_{i\to j}-W_1\right) \\
    & = \mathcal{N}_{1\to i}+W_i-W_1\, ,
\end{aligned}
\end{equation}
where $\vec{x}_i$ labels the position of the region emerging from node $i$.

In practice, when exploring the tree, the value of $V_i$ and $W_i$ is computed on the leaves and propagated upwards as follows.
On the leaves, $W_j=0$, while $V_j=V(L_j)$ is obtained as explained above.
Then, when a node $i$ gives rise to two child nodes $\ell$ and $m$, one has
\begin{equation}
	\label{eq:iter:V:W}
	V_i=V_\ell+V_m
	\quad\text{and}\quad
	W_i =  \frac{1}{V_i} \left[ V_\ell \left( \mathcal{N}_{i\to\ell} + W_\ell\right)+V_m \left( \mathcal{N}_{i\to m} + W_m\right)\right] .
\end{equation}
In this way, using \cref{eq:zeta:N:W}, the curvature perturbation coarse-grained over the regions emerging from each node in the tree can be computed.\footnote{Combining \cref{eq:zeta:N:W,eq:iter:V:W}, one obtains $\zeta_i = (V_\ell \zeta_\ell+V_m\zeta_m)/V_i$, which is of course expected.
	\label{footnote:zetai:zetam:zetal}}

Let us note that, since this propagation proceeds upwards (\ie from the leaves to the primeval patch), 
the value of $W_1$ is computed at the very end only, while it is needed to evaluate the curvature perturbation in \cref{eq:zeta:N:W}.
This implies that the tree has to be explored twice: first to compute $W_i$ and $V_i$, and second to compute $\zeta_i$.
In practice, the recursive embedding of the tree, as detailed in \cref{sec:tree:structure}, does not rely on storing the whole tree once its exploration is complete.
This is one of the reasons why the stochastic-tree procedure is numerically efficient, as it does not require a large amount of dynamic memory.
The tree is not created and then explored; instead, the two processes are simultaneous and intertwined.

In practice, to enable us to explore the same tree realization multiple times, we store the seed from which the Pseudo-Random Number Generator (PRNG) is initiated at the onset of the tree.
As a consequence, each tree realization is determined uniquely by its seed and can be reconstructed at will.
For our PRNG, we use the \texttt{xoshiro256**} pseudorandom number generator included in GFortran.
This generator is thread-safe when using OpenMP and has a period of $2^{256}-1 \sim 10^{77}$.

\subsection{Mapping the curvature perturbation in comoving space}
\label{sec:maps}

\begin{figure}
	\centering
	\begin{subfigure}{0.32\textwidth}
		\includegraphics[width=\textwidth]{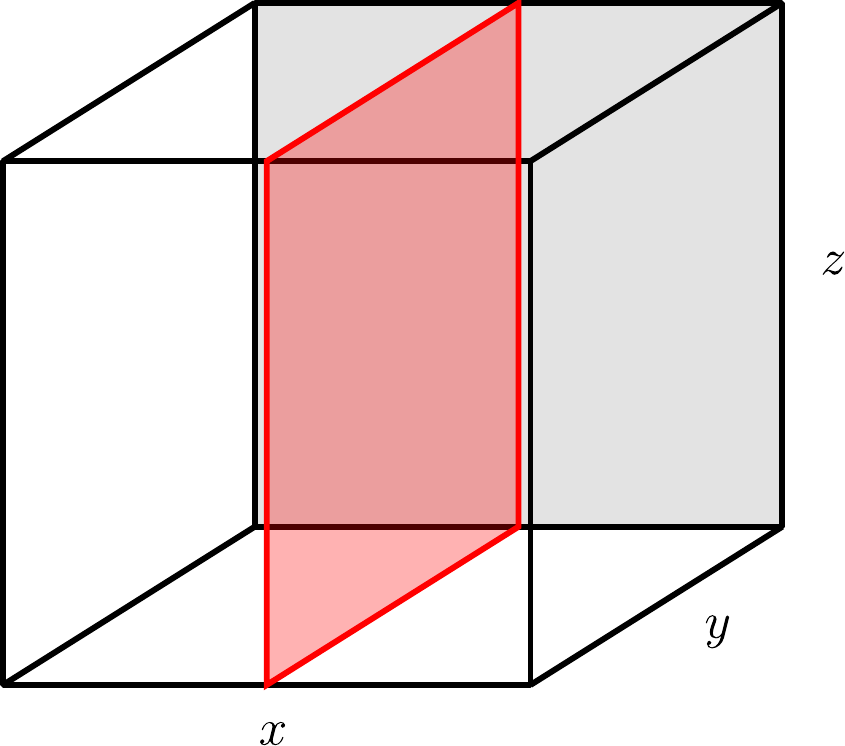}
		\caption{$\Delta N$: we divide the $x$-axis.}
	\end{subfigure}
	\hfill
	\begin{subfigure}{0.32\textwidth}
		\includegraphics[width=\textwidth]{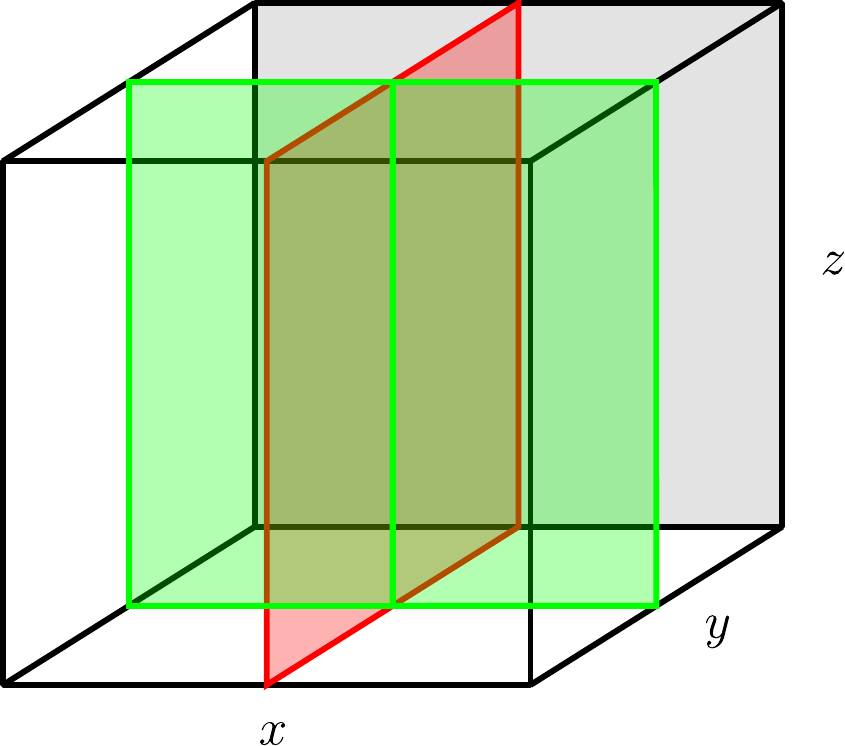}
		\caption{$2 \Delta N$: we divide the $y$-axis.}
	\end{subfigure}
	\hfill
	\begin{subfigure}{0.32\textwidth}
		\includegraphics[width=\textwidth]{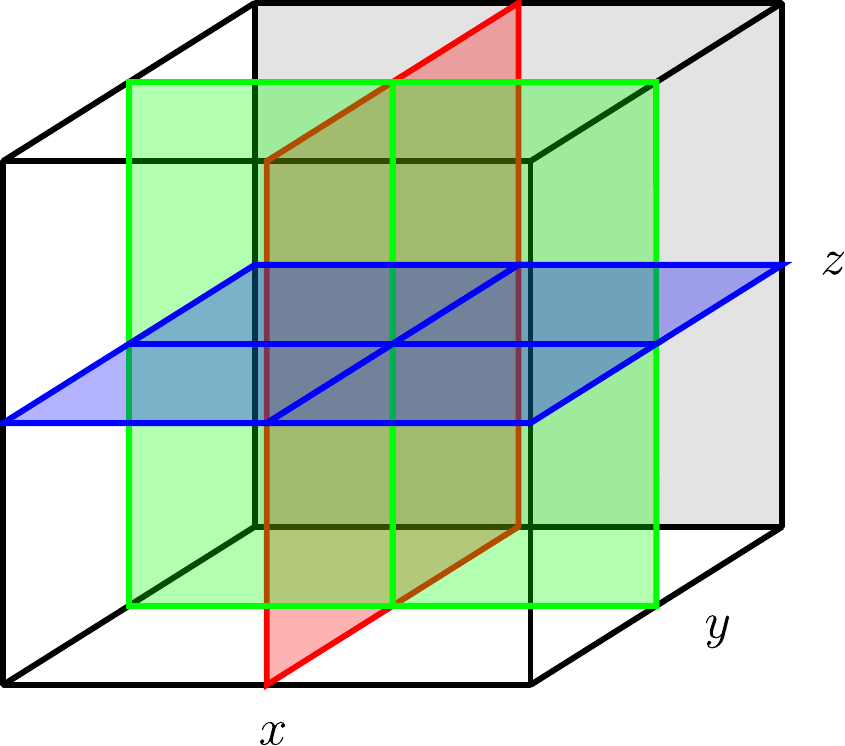}
		\caption{$3 \Delta N$: we divide the $z$-axis.}
	\end{subfigure}
	\caption{Construction of maps in comoving coordinates by recursively dividing growing Hubble patches.}
	\label{fig:map_procedure}
\end{figure}

So far, we have explained how to compute the curvature perturbation at each leaf (\ie within each Hubble patch belonging to the final hypersurface), as well as its value coarse-grained across the descendant leaves of a given node. To derive actual maps of the curvature perturbation, physical coordinates need to be assigned to the leaves, such that their relative position in real space is prescribed. 

In a given tree, there is no unambiguous way to assign physical coordinates to the leaves, or physical distances between them. Consider for instance the one-dimensional tree depicted in \cref{fig:example:tree}, and swap the nodes 10 and 11. Leaves 11 and 6 are not neighbour leaves anymore, and they end up being distant by three (instead of one) Hubble diameters across the end-of-inflation hypersurface. However, by swapping two sibling nodes one does not modify the tree (one is still dealing with the same stochastic realization of the tree process), one only changes its graphical representation. The problem is that, at each branching node, \ie at each volume doubling, two independent branches emerge but the geometry of the regions they describe has been left ambiguous so far. 

This ambiguity needs to be lifted somewhat arbitrarily. 
In practice, we proceed according to the prescription sketched in \cref{fig:map_procedure}. We start from the primeval patch, \ie the root of the tree, pictured as a cube, and we let it evolve until its volume doubles after $\Delta N = \ln(2) / 3$ (or until it reaches the end of inflation). If the patch is still inflating, we split it along the $x$-axis, \ie we produce two Hubble patches separated by an $y-z$ plane. We carry this procedure recursively, and permute at each level of the tree the direction of the splitting as $(xyz)$. In this way, comoving coordinates are assigned to each leaf, corresponding to their position within the cube.

This prescription breaks the permutation symmetry between sibling nodes, which is ultimately related to homogeneity and isotropy of  Friedmann-Lema\^itre-Robertson-Walker spacetimes. This symmetry is broken for each individual tree but at large distances it re-emerges in stochastic averages over tree realizations. This is shown explicitly in \cref{sec:discretization:artefact} below, where we find that standard results  from quantum field theory (QFT) in curved spacetimes for a light test field can be recovered with stochastic trees, and where we further discuss the impact and mitigation of discretization artefacts. 

Note that other spatial discretization schemes may be adopted. For instance, if the orientation of the splitting plane between two sibling Hubble patches is chosen randomly at each step, translational invariance is recovered even at small scales when ensemble average is performed over the trees. However, even in that case it remains true that known QFT results are recovered only at large distances. In fact, in \Refa{Jain:2019gsq} it is found that the convergence towards QFT results is slower than with the na\"ive prescription of \cref{fig:map_procedure}, and that the size of the inflating region at a given time is also not well accounted for (contrary to the prescription of \cref{fig:map_procedure}, which performs very well). Given that the stochastic-$\delta N$ formalism provides approximations for the large scales anyway, we will thus stick with the prescription of \cref{fig:map_procedure}.

\begin{figure}
	\centering
	\begin{subfigure}{0.49\textwidth}
		\includegraphics[width=\textwidth]{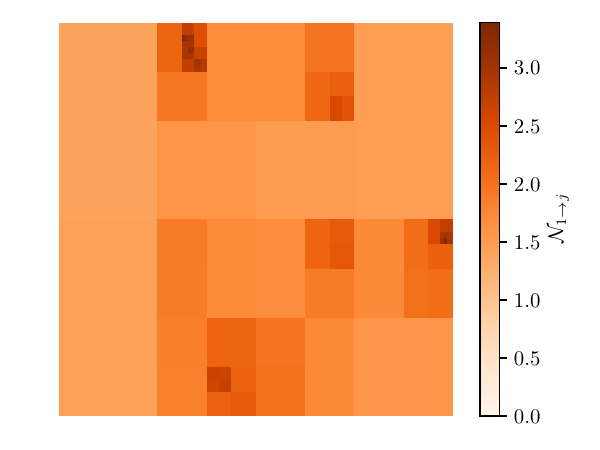}
		\caption{Without PBHs.}
	\end{subfigure}
	\begin{subfigure}{0.49\textwidth}
		\includegraphics[width=\textwidth]{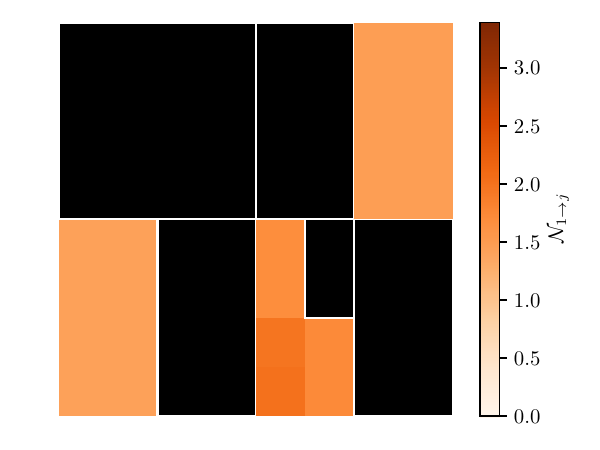}
		\caption{With PBHs.}
	\end{subfigure}
	\caption{
		Example of a two-dimensional comoving map with five PBHs in the flat-well toy model discussed in \cref{sec:application}, with $\mu=0.8$ and starting from $\phi=\Delta\phi$.
		The colour scale indicates the elapsed number of \efolds of the different patches on the end-of-inflation hypersurface.
		This is the same realization as in \cref{fig:tree:example:flat:well}.
		On the right panel, we mark the five PBHs by superimposing black patches, according to the PBH formation criterion presented in \cref{sec:unbalance}.
		We single out this realization for illustrative and pedagogical purposes, 
		it is not particularly representative of the tree population, in which PBH formation remains a rare event.
	}
	\label{fig:cmap}
\end{figure}

In \cref{fig:cmap}, we show the two-dimensional comoving map associated with the tree of \cref{fig:tree:example:flat:well}, where we restrict to two dimensions for the sake of clarity. In practice, we carry even splittings across the $x$-axis -- left child is on the left and right child is on the right -- and the odd splittings across the $y$-axis -- left child is at the top and right child is at the bottom.

Let us stress that the above procedure allows one to construct \emph{comoving} maps, which do not directly display the physical distance between leaves. However, the value of the scale factor within each leaf $j$, as measured with respect to the primeval patch, is nothing but $a(\vec{x}_j)=e^{\mathcal{N}_{1\to j}}$, where $\mathcal{N}_{1\to j}$ is precisely the quantity being displayed in \cref{fig:cmap}. As a consequence the metric $\mathrm{d}\ell^2=a^2(\vec{x}) \mathrm{d} \vec{x}^2$ on the end-of-inflation hypersurface is known, and physical distances across that hypersurface can be computed by solving for geodesics and integrating $\dd{\ell}$ along them. We do not need maps with physical coordinates in the present article and thus leave their analysis for future work.

\section{Harvesting primordial black holes}
\label{sec:pbh}

The formation of primordial black holes usually takes place in regions of large curvature.
However, using the value of the curvature perturbation as a criterion for PBH formation is not always well justified, 
in particular when quantum diffusion plays an important role.
The reason is that, as explained in \cref{sec:zeta}, $\zeta$ is defined with respect to a given observer, 
and relative to a ``background'' that encloses their whole observable universe.
However, gravitational collapse into a black hole is a local process, 
and should not depend on averages performed over regions much larger than the Hubble radius when it takes place.
In other words, $\zeta_i$ in \cref{eq:zeta:N:W} is not a fully local quantity 
since it is affected by the volume produced in regions of the tree that are very distant from node $i$, via their contributions to $W_1$.
Using $\zeta_i$ to decide whether the region emerging from node $i$ collapses into a black hole thus seems problematic.

For this reason, other quantities are often considered for PBH formation criteria, 
such as the comoving density contrast~\cite{Young:2014ana, Musco:2020jjb}, 
or its non-perturbative generalization, the compaction function~\cite{Shibata:1999zs, Musco:2018rwt, Kehagias:2024kgk}.
The reason why the comoving density contrast is better suited than the curvature perturbation is that, 
through Poisson equation, it is proportional to the Laplacian of the curvature perturbation, hence in Fourier space it features an additional $k^2$ factor.
This suppresses large-scale contributions in the coarse-grained density contrast, which is thus a more ``local'' tracer than the curvature perturbation.

\subsection{Coarse-shelled curvature perturbation}
\label{sec:coarse-shelled}

In the stochastic trees, we implement an idea similar to the ``coarse-shelled'' curvature perturbation proposed in \Refa{Tada:2021zzj}.
Consider a given node $i$, giving birth to two nodes $\ell$ and $m$ (this is the situation sketched in \cref{fig:elementary:tree}).
Our goal is to use the node $i$ as a local background for the node $\ell$, 
and to evaluate the curvature perturbation in node $\ell$ relative to its local environment $i$,
\begin{equation}
	\label{eq:zeta_li:def}
	\zeta_{\ell i} = \zeta_\ell-\zeta_i\, .
\end{equation}
This ``coarse-shelled'' curvature perturbation can be related to the Laplacian of the curvature perturbation, or more generally to the compaction function, as follows.
The region emerging from node $i$ is composed of the regions stemming from its two child nodes $\ell$ and $m$, but the geometry of these subregions is not specified.\footnote{
	When constructing the maps of \cref{sec:maps}, a choice was made that consists in splitting volumes along orthogonal planes, 
	the orientation of which alternates as the time elapsed from the primeval patch increases, such that statistical isotropy is recovered at large scales, see \cref{sec:discretization:artefact}. 
It can thus be used to compute quantities coarse-grained over large volumes. The coarse-shelled curvature perturbation being a local quantity, one should rather employ a scheme that is isotropic already at the level of an individual leave and its parent. Note that if the ``concentric-sphere'' scheme were embedded globally it would lead to some Hubble patches being pictured as thin layers of large radius, which would be incompatible with the geometry of inflating space-times. Therefore, different schemes should be employed for local and global quantities, such that the space-time symmetries are correctly implemented at both levels.
}
Since the Laplacian is a spherically symmetric operator, let us picture the regions emerging from nodes $i$ and $\ell$ as being two concentric spheres.
In this case, $\zeta_{\ell i}$ is the curvature perturbation coarse grained over the shell located between the spheres of radius $R_\ell$ and $R_i$, 
where $V_i=4\pi R_i^3/3$ and likewise for $V_\ell$, which is why it is called the ``coarse-shelled'' curvature perturbation.
For the noise appearing in the Langevin equation~\eqref{eq:Langevin} to be white, \ie uncorrelated over time, 
coarse graining has to be performed by means of a sharp window function in Fourier space, $\zeta_R(\vec{k}) = \theta(a/R-k)\zeta(\vec{k})$,
where $\theta(x)$ is the Heaviside function.
In this case,
\begin{equation}
	\zeta_{\ell i}(\vec{k}) = \mathcal{W}_{R_\ell,R_i}(k) \zeta(\vec{k})
	\quad\text{where}\quad
	\mathcal{W}_{R_\ell,R_i}(k)= \theta(k-a/R_i)\theta(a/R_\ell-k) .
\end{equation}
The window function for the coarse-shelled curvature perturbation, $\mathcal{W}_{R_\ell,R_i}(k)$, removes all scales larger than $R_i$, hence it ensures that $\zeta_{\ell i}$ is indeed a local quantity.

A correspondence between $\zeta_{\ell i}$ and the density contrast or the compaction function can be established by matching their effective window functions.
At linear order in perturbation theory, the comoving density contrast is related to the curvature perturbation as
\begin{equation}
	\delta = \frac{2(1+w)}{5+3w} \frac{1}{a^2 H^2}\Delta\zeta\,,
\end{equation}
where $w=p/\rho$ denotes the equation-of-state parameter of the background fluid.
If it is coarse grained with a Gaussian window function~\cite{Young:2019osy, Tokeshi:2020tjq}, $\delta_R(\vec{k})=e^{-(k R/ a)^2/2}\delta(\vec{k})$, 
its relation to the curvature perturbation is given by
\begin{equation}
	H^2 R^2\frac{5+3w}{2(1+w)}
	\delta_R(\vec{k}) = \mathcal{W}_R(k) \zeta(\vec{k})\,,
	\quad\text{where}\quad
	\mathcal{W}_R(k)=  \frac{k^2 R^2}{a^2} \exp[-\frac{1}{2}\left(\frac{k R}{a}\right)^2]\, .
\end{equation}
In practice, PBH formation criteria need to be evaluated at the time when the relevant scale $R$ re-enters the Hubble radius, thus we assume $HR=1$ in the above.

The compaction function $\mathcal{C}(r)$ is defined as the difference between the Misner–Sharp mass contained in the sphere of comoving radius $r$, 
and the background mass within the same areal radius.
Around spherically-symmetric peaks, it is related to the curvature perturbation via
\begin{equation}
	\mathcal{C}(r)=\frac{3(1+w)}{5+3w}\left\{ 1-\left[1+r\zeta'(r)\right]^2\right\} .
\end{equation}
If $r_{\mathrm{m}}$ denotes the value of $r$ where $\mathcal{C}$ is maximum, PBH formation thresholds have been derived on $\mathcal{C}(r_{\mathrm{m}})$, where the
mass of the resultant black hole is related to the one contained within $r_{\mathrm{m}}$.
When the Misner–Sharp mass is defined with a Gaussian smoothing function~\cite{Kalaja:2019uju}, in \Refa{Tada:2021zzj} it is shown that
\begin{equation}
	R\zeta'(R) \simeq - \frac{1}{3}\mathcal{W}_R(k) \zeta(\vec{k})\, ,
\end{equation}
which features the same window function as the comoving density contrast.
A correspondence between the coarse-shelled curvature perturbation, the density contrast and the compaction function can thus be established provided the replacement
\begin{equation}
	\label{eq:W:corr}
	\mathcal{W}_{R_\ell,R_i}(k) \sim \alpha \mathcal{W}_{\beta R_i}(k)
\end{equation}
can be made, where $\alpha$ and $\beta$ should be set such that both hands of \cref{eq:W:corr} are as similar as possible.
In practice, we require that when employed with a logarithmic measure both window functions peak at the same scale 
and that they share the same ``volume'' $\int \mathcal{W}(k) \dd{\ln k}$, which leads to $\alpha=\ln(R_i/R_\ell)$ and $\beta=1$.
As a consequence, a PBH threshold $\delta_{\mathrm{c}}$ on the comoving density contrast can be translated into a threshold $\zeta_{\ell i, \mathrm{c}}$ 
on the coarse-shelled curvature perturbation according to
\begin{equation}
	\zeta_{\ell i, \mathrm{c}} = \frac{5+3 w}{2(1+w)}\ln\left(\frac{R_i}{R_\ell}\right) \delta_{\mathrm{c}}\, .
\end{equation}
Likewise, a PBH threshold $\mathcal{C}_{\mathrm{c}}$ on the compaction function can be translated into
\begin{equation}
	\label{eq:zeta:cs:crit:compaction}
	\zeta_{\ell i, \mathrm{c}} =3 \ln\left(\frac{R_i}{R_\ell}\right)\left[1-\sqrt{1-\left(\frac{5+3w}{3+3w}\right)\mathcal{C}_c}\right] .
\end{equation}

\subsection{Unbalance index}\label{sec:unbalance}

\begin{figure}
	\centering
	\includegraphics[width=\textwidth]{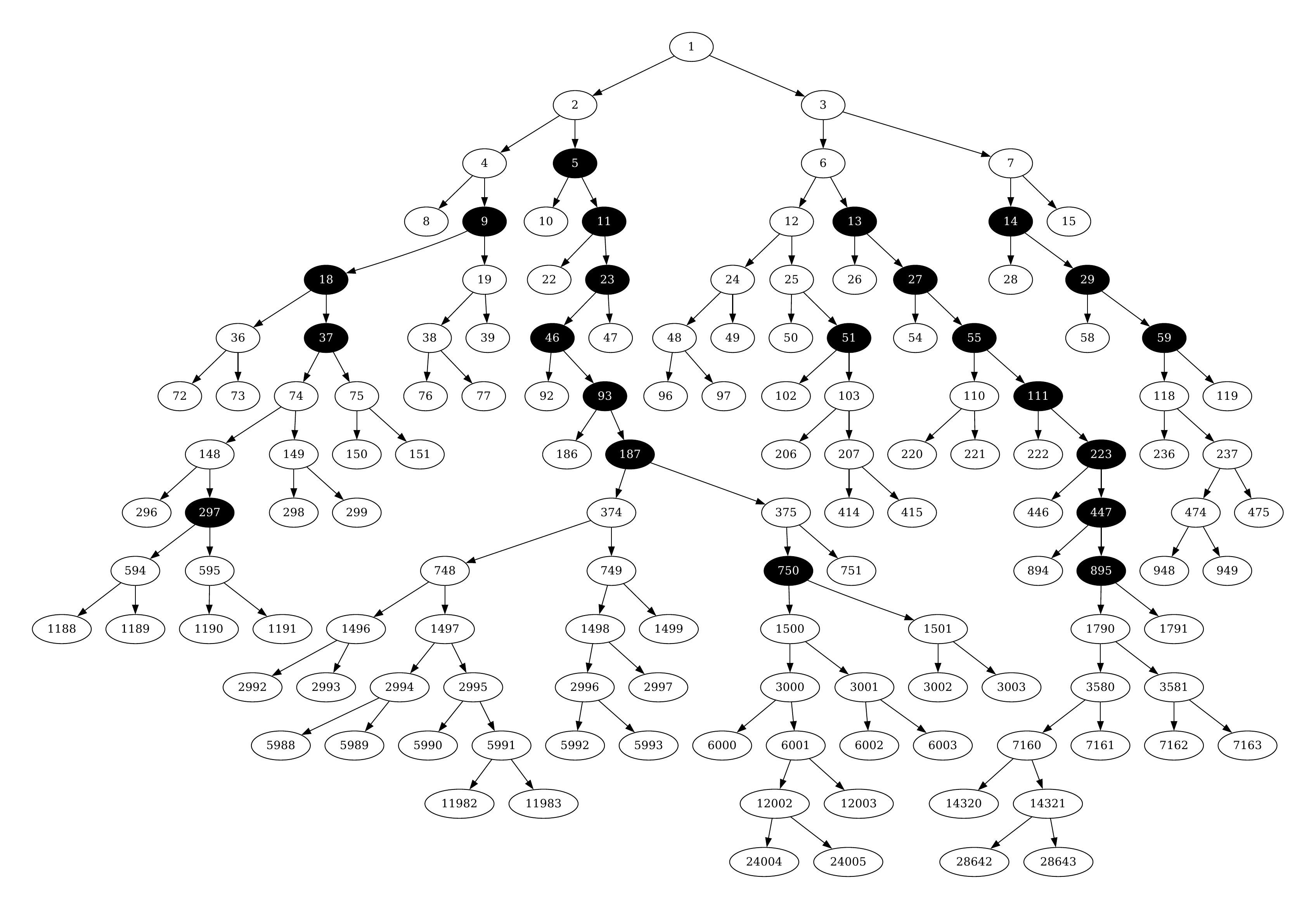}
	\caption{
		Example of a tree realization with five PBHs in the flat-well toy model, with $\mu=0.8$ and starting from $\phi=\Delta\phi$.
		This is the same realization as \cref{fig:cmap}.
		The black patches are those for which the coarse-shelled curvature perturbation, used as a proxy for the compaction function, exceeds the PBH formation threshold.
	}
	\label{fig:tree:example:flat:well}
\end{figure}

In practice, we use \cref{eq:zeta:cs:crit:compaction} to evaluate $\zeta_{\ell i, \mathrm{c}}$ at each branching.
The nodes $\ell$ such that $\zeta_{\ell i}>\zeta_{\ell i, \mathrm{c}}$ are assumed to collapse into PBHs.
Expressed in terms of the compaction function, we take the value $\mathcal{C}_{\mathrm{c}}=0.5$ as a threshold for PBH formation,
assuming that PBHs form in a radiation era with $w=1/3$.\footnote{
	Numerical simulations~\cite{Musco:2018rwt, Germani:2018jgr, Escriva:2019phb, Musco:2020jjb} 
	show that the critical threshold depends sensitively on the details of the density profile around the peak and may vary between $2/5$ and $2/3$.
	Given the approximation made when matching the window functions in \cref{sec:coarse-shelled}, 
	we adopt a fixed value for $\mathcal{C}_{\mathrm{c}}$, 
	on which we have checked that our results do not depend much.
}
This leads to $\zeta_{\ell i, \mathrm{c}} =\ln(V_i/V_\ell)/2$.
In \cref{fig:tree:example:flat:well}, we display a tree realization generated in the flat-well toy model (see \cref{sec:application}), 
where the nodes collapsing to PBHs are coloured in black. 

One can see that the nodes giving rise to PBHs are those for which the two emerging branches generate substantially different volumes.
This can be understood as follows.
By inserting \cref{eq:zeta:N:W} into \cref{eq:zeta_li:def}, one finds
\begin{equation}
	\label{eq:zetali:unbalance}
	\frac{\zeta_{\ell i}}{\zeta_{\ell i, \mathrm{c}}} = \frac{2}{\ln(V_i/V_\ell)} \frac{V_m}{V_i}\left(W_\ell-W_m\right)\,,
\end{equation}
where we have used that $\mathcal{N}_{1\to \ell}=\mathcal{N}_{1\to i}+\mathcal{N}_{i\to \ell}$ together with \cref{eq:iter:V:W}.
We have also assumed that inflation does not end in the nodes $\ell$ and $m$, hence $\mathcal{N}_{i\to \ell}=\mathcal{N}_{i\to m}=\Delta N$. 
If the tree is perfectly balanced between the nodes $\ell$ and $m$, then $W_\ell=W_m$ and $\zeta_{\ell i} = \zeta_{m i} = 0$.
In the opposite limit where the subtree emerging from node $\ell$ is much larger than the one emerging from node $m$, $V_\ell\gg V_m$, 
one can expand $\ln(V_i/V_\ell)\simeq V_m/V_i$, which leads to 
\begin{equation}
	\frac{\zeta_{\ell i}}{\zeta_{\ell i, \mathrm{c}}} \simeq 2(W_\ell-W_m) \,.
	\label{eq:threshold}
\end{equation}
If $V_\ell\gg V_m$, then in most cases $W_\ell\gg W_m$ and the condition $\zeta_{\ell i}>\zeta_{\ell i, \mathrm{c}}$ is met.

The coarse-shelled curvature perturbation of \cref{eq:zetali:unbalance} may thus be seen as an ``unbalance index'':
PBHs form at nodes where the tree exhibits a high level of unbalancing.
This allows one to interpret \cref{fig:tree:example:flat:well} readily, since the black nodes are indeed the most asymmetric ones.

\subsection{Mass distribution}

Once a node giving rise to a PBH has been identified, its mass can be computed.
As a first approximation, it is given by the mass $M_H(R_i)$ comprised in a Hubble volume 
at the time when the scale $R_i$ re-enters the Hubble radius.
Assuming that reheating proceeds instantaneously at the end of inflation, one has
\begin{equation}
	\label{eq:mass:PBH}
	M_H(R_i) \simeq \Mp^2 R_i^2 H_\mathrm{end}\, .
\end{equation}
A more precise estimate can be obtained accounting for critical collapse~\cite{Niemeyer:1997mt}, 
but the associated scaling law is valid only when $\mathcal{C}-\mathcal{C}_{\mathrm{c}}\lesssim \mathcal{O}(10^{-2})$~\cite{Kalaja:2019uju}.
Since most super-critical nodes are found to exceed the threshold by more than $10^{-2}$, 
critical scaling is not more accurate than the simple estimate~\eqref{eq:mass:PBH}, to which we thus stick.

\subsection{Cloud-in-cloud}
\label{subsec:cloud:in:cloud}

As can be seen in the example displayed in \cref{fig:tree:example:flat:well}, when a subtree deviates from the average trajectory and produces a large volume, it induces a substantial unbalancing not only of the node it starts from, but also of some of its parent nodes. The reason is that all parent nodes give rise to one branch that holds this subtree and one branch that does not, hence they are likely unbalanced by its large volume. As a consequence, it is common that PBHs arise at multiple nodes along a path segment, hence that they form in a nested way. 

Nested PBH formation is analogous to the well-known cloud-in-cloud problem~\cite{Jedamzik:1994nr}, in which a hierarchy of overdense regions leads to PBH formation inside larger PBHs.
Fortunately, the recursive structure of stochastic trees is perfectly suited to treat this issue.
Indeed, in \texttt{FOREST}, PBH formation is evaluated dynamically at each node while the tree is created.
Information such as $V_i, W_i$, the number and masses of PBHs is always propagated upwards and saved in the stack as we progress in the tree structure.
As a consequence, if the algorithm finds a new PBH at a node, then all its inner PBHs are discarded and do not propagate further. ``Cloud-in-cloud'' is therefore automatically accounted for in this approach.

\section{Application: quantum-well model}
\label{sec:application}

As a direct application of the framework we presented and to showcase the power of \texttt{FOREST}, we revisit the common toy model of quantum-well inflation~\cite{Pattison:2017mbe, Ezquiaga:2019ftu, Tada:2021zzj, Ando:2020fjm, Animali:2024jiz}.
In this framework, the inflaton field leaves its classical trajectory, dominated by the potential-induced drift, when it enters a flat region of the potential $V(\phi) = 24 \pi^2 \Mp^2 v_0$, where the dynamics is driven by the stochastic noise.

We assume that the flat well ends at $\phi_\mathrm{end}$\footnote{
    The field value $\phi_{\mathrm{end}}$ marks with the end of inflation in our setting. 
    If the exit from the well does not coincide with the end of inflation, but instead a classical phase follows, the number of inflationary \efolds would be shifted by a constant value. 
    However, this shift does not affect the amplitude of the curvature perturbations, although it does shift the scales and the masses of collapsed objects, by a constant multiplicative factor.
}, which thus features an absorbing boundary for our Langevin trajectories.
The quantum well also exhibits a reflective boundary at $\phi_\mathrm{end} + \Delta \phi$, with $\Delta\phi$ being the width of the well,
separating the region of quantum diffusion from the region of classical evolution prior to it. 

It is convenient to parametrize the model in terms of the following dimensionless quantities
\begin{equation}
	\mu^2 \equiv \frac{\Delta \phi^2}{v_0 \Mp^2}  \quad \text{and} \quad 
	x \equiv \frac{\phi - \phi_\mathrm{end}}{\Delta \phi} \in [0, 1]\,.
\end{equation}
Qualitatively, $\mu$ is an indicator of the time scales associated to the flat well. For instance, the mean number of \efolds across the well is given by~\cite{Pattison:2017mbe} $\langle \mathcal{N}  \rangle = \mu^2/2$. One can also show that the mean volume emerging from a patch with field value $x$, in Hubble units, is given by~\cite{Animali:2024jiz}
\begin{equation}\label{eq:mean:vol:flat}
 \ev{V}= \ev{e^{3 \mathcal{N}}}  =\frac{\cos{[\sqrt{3}\mu (1-x)]}}{\cos{(\sqrt{3}\mu)}}\,,
\end{equation}
which is well-defined only up to a critical value $\mu<\mu_{\mathrm{c}} \equiv \pi/(2\sqrt{3})$.\footnote{
    The first equality $\ev{V} = \ev{e^{3\mathcal{N}}}$ is not trivial but can be proven rigorously using branching processes.}
For values of $\mu \geq \mu_{\mathrm{c}}$, the tail of the first-passage-time distribution decays more slowly than $e^{3\mathcal{N}}$, and the mean volume is divergent, which may lead to an ``eternal inflation'' problem (see discussion in \cref{subsec:vol:weight}).
For this reason, we only work with values of $\mu$ below the critical value.

\subsection{Probability distributions over the trees}
\label{subsec:tree:populations}

\begin{figure}
	\begin{subfigure}[t]{.49\textwidth}
		\includegraphics[width=\textwidth]{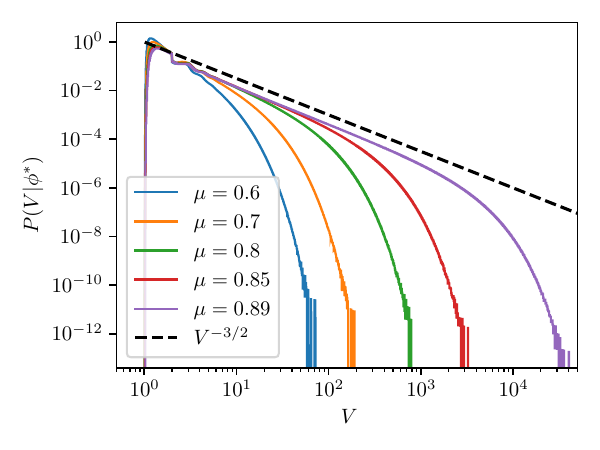}
		\caption{Probability distribution of the final volume.}
	\end{subfigure}
	\begin{subfigure}[t]{.49\textwidth}
		\includegraphics[width=\textwidth]{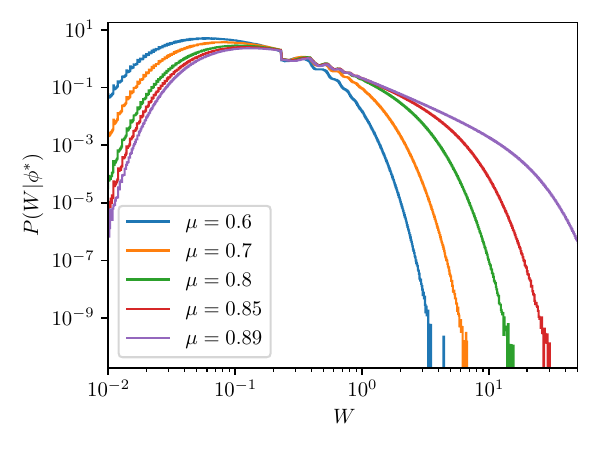}
		\caption{Probability distribution of the volume-averaged expansion.}
	\end{subfigure}
	\caption{
		Probability distribution of the final volume (a) and of the volume-averaged number of \efolds (b) over the tree population in the flat well for $\mu = 0.6, 0.7, 0.8, 0.85, 0.89$ and $x_*=1$.
		These were obtained with, respectively, $10^{11}, 10^{11}, 5\times 10^{10}, 2\times 10^{10}, 10^{10}$ trees.
	}
	\label{fig:flat-pop}
\end{figure}

Starting from a patch with field configuration $x_* \in [0, 1]$, we let it grow for $\Delta N = \ln(2) / 3$ then split it recursively.
This produces a tree configuration of which we compute the volume $V$ and the volume-averaged number of \efolds $W$.
By simulating a population of such trees, we are able to sample the distributions of $V$ and $W$ over the ensemble of trees.
We show these distributions for the flat-well model and for several values of $\mu$ in \cref{fig:flat-pop}.

First, we observe that the tree populations have a modest mean volume $\ev{V} = \order{10}$, as well as a modest mean value for the volume-averaged expansion $\ev{W} = \order{1}$. 
Both distributions, however, exhibit exponential tails, which become progressively heavier as $\mu$ increases. 
This is to be expected, given that quantum-diffusion effects are larger for increasing values of $\mu$. 
In particular, the volume distribution shows the behaviour $P(V|\phi_*)\propto e^{z_* V} V^{-3/2}$, where $z_* < 0$ is the rightmost pole of the generating function for $V$~\cite{Winitzki:2008ph}.
Above its mean value, the volume distribution follows a power-law behaviour, which is then overtaken by an exponential decay at the far tail. 

Increasing $\mu$ shifts the onset of this exponential phase to larger volumes, resulting in heavier tails. 
For $\mu \to \mu_{\mathrm{c}}$, $z_* \to 0$, which implies $P(V)\propto V^{-3/2}$: this marks the onset of the eternal-inflation regime, which is represented by the black dashed line in the left panel of \cref{fig:flat-pop}. 
This is in agreement with previous studies in which a similar behaviour for the volume distribution at the end of inflation has been found~\cite{Dubovsky:2008rf, Winitzki:2008ph}. Also note that the slight kinks when $V$ and $W$ are of order one are discretization artefacts, see \cref{sec:branching:time}, which do not affect the large-scale (\ie large $V$ or large $W$) statistics.

Finally, let us stress that exponential tails are one of the key features of the stochastic-$\delta N$ formalism~\cite{Pattison:2017mbe, Ezquiaga:2019ftu, Vennin:2020kng, Animali:2022otk, Briaud:2023eae, Animali:2024jiz}.
This is of particular interest in the context of PBH formation 
since the relative abundance of anomalously large subtrees is what determines the fraction of the end-of-inflation hypersurface that eventually collapses into PBHs.

\subsection{Probability distributions over the leaves}

\begin{figure}
	\centering
	\includegraphics[width=.7\textwidth]{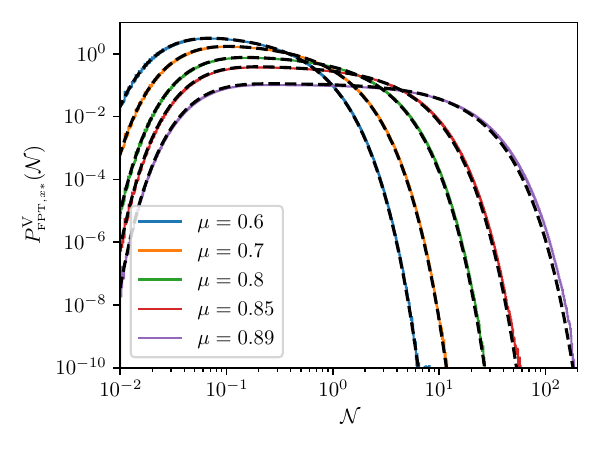}
	\caption{
		Volume-weighted probability distribution of the first-passage time $\mathcal{N}_{x_*\to x_\mathrm{end}}$ through the end-of-inflation hypersurface, in the flat well for $\mu = 0.6, 0.7, 0.8, 0.85, 0.89$ and $x_*=1$.
		These were obtained with, respectively, $10^{11}, 10^{11}, 5\times 10^{10}, 2\times 10^{10}, 10^{10}$ trees. Full coloured lines stand for the results of \texttt{FOREST}, whereas black dashed lines represent the analytical formula~\eqref{eq:FPTVolWeight}.}
	\label{Fig:PZetaCompFlat}
\end{figure}

In addition to global properties of the trees, one may be interested in properties of their leaves, \ie compute statistics over the set of leaves of the entire tree population.
For instance, starting from the root patch $1$ having field configuration $x_*$, one can register for each leaf $j$ its number of \efolds from the root, weighted by its final volume.
This gives rise to the volume-weighted first-passage-time distribution through the end-of-inflation hypersurface, $\Pfpt{x_*}^V(\mathcal{N})$, which is computed with \texttt{FOREST} and shown in \cref{Fig:PZetaCompFlat} for several values of $\mu$ and with $x_*=1$.

We find that $\Pfpt{x_*}^V(\mathcal{N})$ displays a transient power-law  behaviour followed by a heavy exponential tail, showing that a relatively abundant number of branches expands for a very long time compared to the average.
Note that PBH formation is generically dominated by these outliers in the tail of the distribution.

In principle, from the first-passage-time distribution one can also derive the distribution of the curvature perturbation coarse-grained at the scale $R_\sigma$ at the end of inflation, $\zeta_{R_\sigma}$. 
Indeed, the two quantities differ only by the subtraction of the volume-averaged expansion $W_1$ in \cref{eq:zeta:end:tree}. However, as mentioned above, $W_1$ is to be averaged over the region of the final hypersurface a given observer has access to. That region is different for every tree, hence different values of $W_1$ need to be subtracted from leaves belonging to different trees, and the distribution for $\zeta_{R_\sigma}$ is not simply a shifted version of the distribution for $\mathcal{N}$. Since this issue does not arise for the coarse-shelled curvature perturbation, which is relevant for PBH formation that we next study, we limit ourselves to presenting the distribution of $\mathcal{N}$.

\subsection{Distribution of primordial black holes}

\begin{figure}
	\begin{subfigure}[t]{.49\textwidth}
		\includegraphics[width=\textwidth]{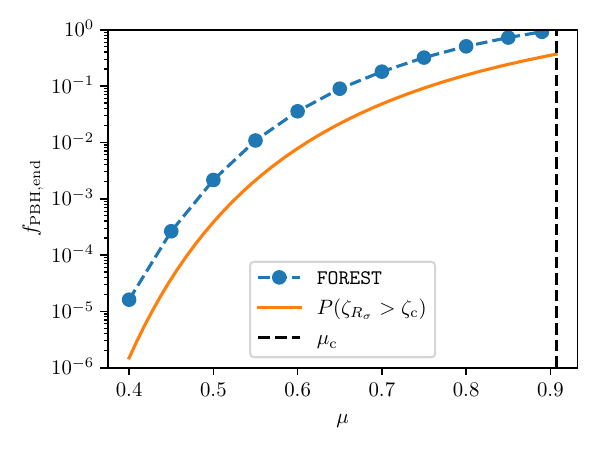}
		\caption{Mass fraction of PBHs.}
		\label{Fig:CompProbxstar0}
	\end{subfigure}
	\begin{subfigure}[t]{.49\textwidth}
		\includegraphics[width=\textwidth]{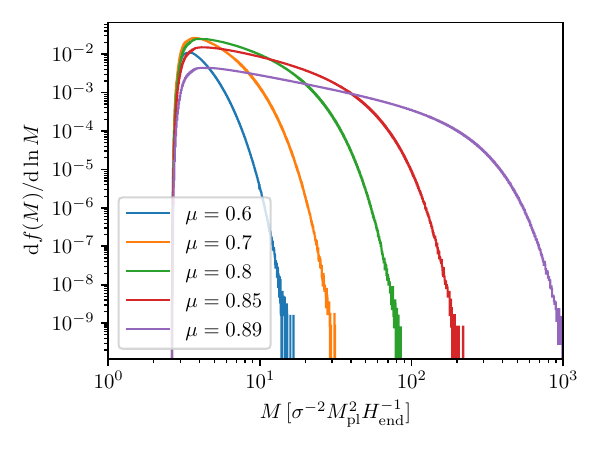}
		\caption{PBH mass distribution.}
		\label{fig:mass-pbh}
	\end{subfigure}
	\caption{
		Statistics of PBH production in the flat quantum well.
		Panel (a): fraction of the universe comprised in (regions that will eventually collapse into) PBHs at the end of inflation, as a function of $\mu$, compared to the simple estimate $P(\zeta_{R_\sigma}>\zeta_{\mathrm{c}})$ derived in the large-volume approximation in \cref{sec:large}. 
		Panel (b): distribution of masses for the produced PBHs for $\mu = 0.6, 0.7, 0.8, 0.85, 0.89$ and $x_*=1$.
		These were obtained with, respectively, $10^{11}, 10^{11}, 5\times 10^{10}, 2\times 10^{10}, 10^{10}$ trees.
	}
	\label{fig:flat-pbhs}
\end{figure}

As described in \cref{sec:coarse-shelled,sec:unbalance}, \texttt{FOREST} identifies and records all the PBHs that appear in our populations of stochastic trees using an ``unbalance index'', \ie a proxy for the compaction function.
As per the name suggests, the unbalance index picks up nodes from which two sibling branches experience very different expansions, which indicates strong gradients in the curvature perturbation.
When we explore our stochastic trees, we only keep track of the utmost/largest PBHs in the tree (see \cref{subsec:cloud:in:cloud}) and we assume that the downstream branch completely collapses, thus setting the mass of the resulting PBH.
This is illustrated in~\cref{fig:tree:example:flat:well,fig:cmap}, which display a tree with five distinct PBHs of relatively large masses.
When averaged over a large ensemble of trees, we report our findings for the flat well and $\mu = 0.6, 0.7, 0.8, 0.85, 0.89$ in \cref{fig:flat-pbhs}.

First, in \cref{Fig:CompProbxstar0} we show $f_\mathrm{PBH,end}$, the fraction of the universe at the end of inflation that will eventually collapse into PBHs, as a function of $\mu$.
We recover the expected result that PBH production is more abundant when quantum diffusion is more prominent, that is when $\mu$ increases.
As $\mu$ gets closer to the critical value $\mu_{\mathrm{c}}$, nearly all the universe is contained in PBHs.

At a first glance, one may be surprised to have $f_\mathrm{PBH,end} \to 1$ close to the critical point, when our criterion for PBH requires having contiguous regions of short and long expansion.
This can be understood by the fact that the overall volume of the universe becomes dominated by a limited number of very large branches leading to PBHs, separated by numerous regions of short expansion whose volume is exponentially suppressed close to the critical value.

Second, we show the mass distribution of these PBHs in \cref{fig:mass-pbh}. We find that the distributions peak at a value of the mass that increases with $\mu$, that they are then followed by a power-law range whose width also increases with $\mu$, before being interrupted by an exponential tail. The existence of these wide and mild power laws is a direct consequence of the ``cloud-in-cloud'' mechanism described in \cref{subsec:cloud:in:cloud}, by which smaller PBHs are concealed within larger ones.

When $\mu$ approaches $\mu_{\mathrm{c}}$, the power-law behaviour extends to larger and larger masses, with $\dv*{f}{\ln M} \propto M^{-\alpha}$, with $\alpha \approx 2/3$.
This can be seen clearly in the case $\mu=0.89$, which is indeed close to criticality. 
PBH formation close to $\mu_\mathrm{c}$ thus displays some properties one expects from critical phenomena: PBH form at all scales and cover nearly all the space in the universe, only separated by infinitely small regions of void.

\section{Discussion}
\label{sec:discussion}

\subsection{Volume weighting}\label{subsec:vol:weight}

Previous attempts to analytically describe the curvature perturbation and PBH formation in stochastic inflation are based on first-passage-time analysis. In this section we review why first-passage-time distributions need to be volume weighted, and we check that they may account for some (but not all) of the results presented above.

In the stochastic-inflation picture, distinct regions of the universe inflate by different amounts, hence they give rise to different numbers of Hubble patches at the end of inflation and contribute differently to ensemble averages over these patches. These ensemble averages correspond to observable quantities: if a hypothetical observer located on the end-of-inflation hypersurface is tasked with computing the average expansion, they will measure the expansion in each patch contained within their observable region and then perform an average over the ensemble of such measurements. Averaging over the set of final Hubble patches, \ie over the set of leaves, is nothing but volume averaging over the set of trees or subtrees, \ie weigh each subtree by the volume it generates~\cite{Animali:2024jiz, Vennin:2024yzl}. In other words, spatial averaging in physical coordinates is equivalent to volume-weighted spatial averaging in comoving coordinates. 

For the first-passage-time distribution associated to the Langevin problem~\eqref{eq:Langevin}, the volume-weighted version can be defined as~\cite{Animali:2024jiz}
\begin{equation}\label{eq:FPTVolWeight}
    \Pfpt{\phi}^{\mathrm{V}}(\mathcal{N})
    =\frac{\Pfpt{\phi}(\mathcal{N}) e^{3\mathcal{N}}}{\int_0^\infty \Pfpt{\phi}(\mathcal{N}) e^{3\mathcal{N}} \dd{\mathcal{N}}}\,,
\end{equation}
which is well-defined only when the integral in the denominator converges. 
If the first-passage-time distribution features an exponential tail $\Pfpt{\phi}(\mathcal{N}) \propto  e^{-\Lambda_0 \mathcal{N}}$ at large $\mathcal{N}$, then it does not converge when $\Lambda_0 < 3$, giving rise to eternal inflation. In that case the mean tree volume is always infinite, regardless of the initial condition.

For the quantum-well model considered in \cref{sec:application}, the first-passage-time problem has an exact solution in terms of the derivative of the second elliptic theta function~\cite{Pattison:2017mbe}, $\Pfpt{x_*}(\mathcal{N})=-\pi/(2 \mu^2)\vartheta_2^\prime(x_* \pi/2, e^{-\pi^2/\mu^2 \mathcal{N}})$. 
Its tail is of the exponential type, $\Pfpt{x_*}\propto e^{-\pi^2/(4 \mu^2)\mathcal{N}}$, which gives rise to the critical value $\mu < \mu_{\mathrm{c}} \equiv \pi/(2 \sqrt{3})$ mentioned below \cref{eq:mean:vol:flat}.
The volume-weighted version \eqref{eq:FPTVolWeight} of this first-passage-time distribution is shown in \cref{Fig:PZetaCompFlat}, where it is compared with the distribution reconstructed from stochastic-tree simulations. We observe an excellent agreement between the two, for all values of $\mu$ scanned and for any value of $\mathcal{N}$.

The reason why this agreement is not so obvious is that, in \cref{Fig:PZetaCompFlat}, the leaves over which the first-passage-time distribution is computed are not the end-point of independent stochastic realizations of the Langevin equation. Indeed, two leaves belonging to the same tree share the same path until their splitting node and thus give rise to values of $\mathcal{N}$ that are correlated. As a consequence, the distribution displayed in \cref{Fig:PZetaCompFlat} is not obtained by sampling independent realizations of the Langevin process: realizations are gathered in trees within which they may be highly correlated. However, when the number of trees is large, these correlations become sparse since the fraction of the leaves to which a given leaf is correlated becomes arbitrarily small (it cannot exceed the ratio between the volume of the tree it lies in and the total volume of all trees). In this limit, leaves are mostly independent and the mere volume weighting ~\eqref{eq:FPTVolWeight} coincides with the ensemble average over leaves. 

This serves both as a robustness test for the numerical code and, more importantly, as a consistency check for the volume-weighting procedure~\eqref{eq:FPTVolWeight}, at least as far as the duration of inflation is concerned. This is important since such procedures are ubiquitous when designing analytical approximations in terms of first-passage-time distributions, such as the large-volume approximation that we now discuss.

\subsection{Large-volume approximation}\label{sec:large}

As mentioned in \cref{sec:intro}, the stochastic evolution blurs the classical relationship between a comoving scale and the value of the inflaton field when that scale crosses out the Hubble radius. 
This led to the development of different approximation schemes. 

The framework that most closely aligns with the idea of stochastic trees is the one presented in~\Refa{Animali:2024jiz}, which relies on the so-called ``large-volume approximation'' that can be summarized as follows. 
Let us consider a region $\mathcal{B}$ of radius $R$ within the end-of-inflation hypersurface, whose parent patch during inflation is denoted as $\mathcal{P}_*$. Field values in the latter are denoted $\vb{\Phi}_*$, where the bold notation indicates that several fields may be at play. 
The physical volume of that region is given by
\begin{equation}
    V=\frac{4}{3}\pi R^3
    = \int_{\mathcal{P}_*} e^{3 \mathcal{N}_{\mathcal{P}_*}(\vec{x})} \dd[3]\vec{x}\,,
\end{equation}
where $\vec{x}$ denotes the comoving coordinate of each spatial point inside the patch $\mathcal{P}_*$, and $\mathcal{N}_{\mathcal{P}_*}(\vec{x})$ is the first-passage time of the worldline attached to $\vec{x}$ through the end-of-inflation hypersurface.

If $R\gg (\sigma H_{\mathrm{end}})^{-1}$, the volume of $\mathcal{B}$ is much larger than the volume of the parent patch $V_*=4/3 \pi [\sigma H(\mathbf{\Phi}_*)]^{-3}$, which, in the language of stochastic trees, corresponds to considering configurations in which a certain parent node generates a large number of leaves.
In this case, according to the central limit theorem, one may replace $e^{3 \mathcal{N}(\vec{x})}$ by its mean value
\begin{equation}
\label{eq:CLT:largeV}
    e^{3\mathcal{N}(\vec{x})}\rightarrow \ev{e^{3\mathcal{N}_{\mathbf{\Phi}_*}}}
    =\int \Pfpt{\mathbf{\Phi}_*}(\mathcal{N}) e^{3\mathcal{N}}\dd{\mathcal{N}} \,,
\end{equation}
where $\Pfpt{\mathbf{\Phi}_*}$ denotes the first-passage time distribution  with initial condition $\vb{\Phi}_*$\,.

As a result, the volume distribution takes the form
\begin{equation}
	P(V| \vb{\Phi}_*) \simeq \delta_{\mathrm{D}}(V-V_* \ev{e^{3 \mathcal{N}_{ \vb{\Phi}_*}}}),
\end{equation}
where $\delta_{\mathrm{D}}$ is the Dirac distribution, and the same holds for the volume-averaged expansion
\begin{equation}
    W=\frac{V_*}{V}\int_{\mathcal{P}_*} e^{3\mathcal{N}_{\mathcal{P}_*}(\vec{x})} \mathcal{N}_{\mathcal{P}_*}(\vec{x}) \dd[3]{x} 
    \rightarrow \frac{V_*}{V} \ev{\mathcal{N}_{\mathbf{\Phi}_*} e^{3\mathcal{N}_{\mathbf{\Phi}_*}}} \,.
\end{equation}

This allows one to establish a correspondence between field values during inflation and physical scales at the end-of-inflation hypersurface, since $V_*\ev{e^{3\mathcal{N}_{\mathbf{\Phi}_*}}} =V\propto R^3$. Further considering
single-clock models of inflation, hypersurfaces of constant forward volume reduce to single points in field phase space, and therefore backward-field values become deterministic quantities.

The large-volume approximation thus differs from the backward approximation~\cite{Ando:2020fjm, Tada:2021zzj}, where a single representative trajectory ending on the region $\mathcal{B}$ is considered to approximate the spatial average of the amount of expansion within that region, and which gives rise to a probability distribution for backward fields. They may be seen as approximations in opposite regimes.

Within the large-volume approximation framework, and in single-clock models of inflation like the one considered in this work, \cref{eq:zeta:N:W} can be approximated as 
\begin{equation}\label{eq:zetaR:LVA}
	\zeta_R \simeq \mathcal{N}_{\phi_0\rightarrow \phi_*}
	+\ev{\mathcal{N}_{\phi_*}}_{\mathrm{V}}-\ev{\mathcal{N}_{\phi_0}}_{\mathrm{V}}\,,
\end{equation}
where $\phi_0$, $\phi_*$ are the inflaton values at nodes $1$ and $i$ respectively and $R$ is the physical radius of the region emerging from node $i$.
In the above expression, the first term denotes a first-passage time, while the latter two terms are volume-weighted stochastic averages, 
$\ev{\mathcal{N}_\phi}_{\mathrm{V}}=\int_0^\infty \dd{\mathcal{N}} \mathcal{N} \PfptV{\phi}(\mathcal{N}).$
This ultimately leads to expressing the one-point distribution of the curvature perturbation, coarse-grained over a region of size
$R$ on the end-of-inflation hypersurface, as
\begin{equation}\label{eq:LVA:P:zeta}
	P(\zeta_R)
	=\PfptV{\phi_0\rightarrow\phi_*}\left(\zeta_R-\ev{\mathcal{N}_{\phi_*}}_{\mathrm{V}}
	+\ev{\mathcal{N}_{\phi_0}}_{\mathrm{V}}\right) .
\end{equation}

Let us now see how predictions of the large-volume approximation compare with the results from stochastic trees, focusing on quantities relevant for the study of primordial black holes.

Although, as explained in \cref{sec:pbh}, the curvature perturbation is not the optimal cosmological field when it comes to discussing PBHs, let us consider it for the sake of illustration, and let us assume that black holes form when it exceeds a certain critical value, if coarse grained over a region of a certain size $R$.
In the Press-Schechter approximation~\cite{Press:1973iz}, the abundance of PBHs is determined by the integral
\begin{equation}\label{eq:above:thresh:prob}
	P[\zeta_R>\zeta_{\mathrm{c}}]=\int_{\zeta_{\mathrm{c}}}^\infty P(\zeta_R) \dd{\zeta_R}\,,
\end{equation}
where $\zeta_{\mathrm{c}}$ is the PBH formation threshold, here assumed $\zeta_{\mathrm{c}}=1$\,.
The above is the probability of a certain region of size $R$ to lie above the threshold, or equivalently, the probability to collapse into a structure of size at least $R$.
The large-volume approximation provides access to the probability~\eqref{eq:above:thresh:prob}, from which the mass fraction and mass distribution of PBHs can thus be derived.

However, this approximation is valid only when certain conditions are met. The main requirement is that large volumes are generated, in order for the central-limit replacement~\eqref{eq:CLT:largeV} to be performed. 
In other words, the mean volume emerging from a given backward field value $\phi_*$ must be much larger than one (in Hubble units). 
Yet, as shown in the left panel of \cref{fig:flat-pop}, see also \cref{eq:mean:vol:flat}, for values of $\mu\leq 0.89$ considered with \texttt{FOREST} the mean volume is only of order a few. This implies that the large-volume approximation may not provide a reliable description of the model considered here.

In fact, even in models yielding large volumes, doubts may be cast regarding the ability of the large-volume approximation to describe extreme events such as PBH formation. The reason is that, as mentioned above, PBHs form when a subtree grows anomalously large. This can be readily seen in \cref{fig:mass-pbh}, where most PBH masses exceed by far the maximal mean volume (respectively given by $2.0$, $2.9$, $5.4$, $10.2$ and $34.2$ for the values of $\mu$ displayed in \cref{fig:mass-pbh}, in increasing order, see \cref{eq:mean:vol:flat}). In this case, approximating volumes by their mean value is incorrect, and the replacement~\eqref{eq:CLT:largeV} is not justified.  PBHs remain outlier events living in the tail of the distribution, and cannot be described directly with a central limit theorem.

As a consequence, here we do not compare directly the mass distributions, but instead we focus on the abundance of PBHs at the end of inflation. 
In practice, this corresponds to the limiting case where $\phi_*\rightarrow \phi_{\mathrm{end}}$, such that \cref{eq:zetaR:LVA} becomes $\zeta_{R_{\mathrm{end}}}\simeq \mathcal{N}_{\phi_0\rightarrow\phi_{\mathrm{end}}}-\langle \mathcal{N}_{\phi_0}\rangle_{\mathrm{V}}$, where $R_{\mathrm{end}}=(\sigma H_{\mathrm{end}})^{-1}$, and we are coarse-graining in a single leaf. 
In the flat-well model, this leads to the one-point distribution~\cite{Animali:2022otk}
\begin{equation}
    P(\zeta_{R_{\mathrm{end}}})=
    -\frac{\pi}{2 \mu^2} \cos(\sqrt{3}\mu)
    \vartheta_2^\prime \left\{
        \frac{\pi}{2},
        e^{-\frac{\pi^2}{ \mu^2}\left[\zeta_{R_{\mathrm{end}}}+\frac{1}{2\sqrt{3}}\mu \tan{(\sqrt{3}\mu)}\right]}
    \right\}
    e^{3 \zeta_{R_{\mathrm{end}}}+\frac{\sqrt{3}}{2}\mu \tan{(\sqrt{3}\mu)}}\,,
\end{equation}
and the corresponding mass fraction, according to \cref{eq:above:thresh:prob}, is shown as a function of $\mu$ as the orange line in \cref{Fig:CompProbxstar0}.

This estimate exhibits the same behaviour as the full result obtained with \texttt{FOREST}, and is even in quantitative agreement to within about one order of magnitude.
This is surprising, given the significant differences between the two approaches: \texttt{FOREST} addresses the cloud-in-cloud problem, which is not considered in the large-volume approximation. 
Moreover, the latter employs the curvature perturbation, rather than the coarse-shelled curvature perturbation, as the criterion for PBH formation. 
Finally, through \cref{eq:zetaR:LVA}, the large-volume approximation also provides a prescription for the averaged expansion to subtract, namely $W_1 \simeq \langle \mathcal{N}_{\phi_0}\rangle_{\mathrm{V}}$, though, as discussed above, in principle this is a non-local, tree-dependent quantity. Interestingly, this implies that qualitative estimates of the mass fraction generally made using the proxy \eqref{eq:above:thresh:prob} in the literature prove to be fairly reasonable estimates. They would fail to predict the details of the mass distributions (or even the typical masses at which PBHs form), and presumably other quantities such as the typical distances over which PBHs are correlated, but for their overall abundance they seem reliable.

\subsection{Discretization artefacts}
\label{sec:discretization:artefact}

In the framework introduced above, space is described as a tessellation of Hubble patches, and branches split in half at fixed time intervals $\Delta N$.
Stochastic trees are thus objects that are discrete both over space and time.
This implies that spacetime properties below and around the Hubble scale are not properly captured by stochastic trees, which can only provide insight into the structure of inflating spacetimes over long times and large distances.
In this section, we discuss artefacts coming from the discretization procedure, check that they become indeed suppressed at large scales and show that we can reproduce known results from quantum field theory in de Sitter.

For the sake of simplicity, we focus on a light test scalar field in a fixed binary tree.\footnote{Instead of a binary tree, one may consider trees with more branches attached to each node (for instance, trees where Hubble patches split into $8$ patches after $\Delta N=\ln(2)$ \efolds in $3$ dimensions). This would not change the nature of the arguments developed here.}
The field being test, inflation ends at the same time in all the branches, \ie the tree is perfectly balanced.
This makes it easier to relate topological distances in the tree and physical distances on the final hypersurface. Although it may be seen as restrictive, it will allow us to show explicitly how discretization artefacts disappear at large scales, and to argue why this is in fact a generic result.

\subsubsection{Branching times}
\label{sec:branching:time}

\begin{figure}
	\centering
	\includegraphics[width=0.3\textwidth]{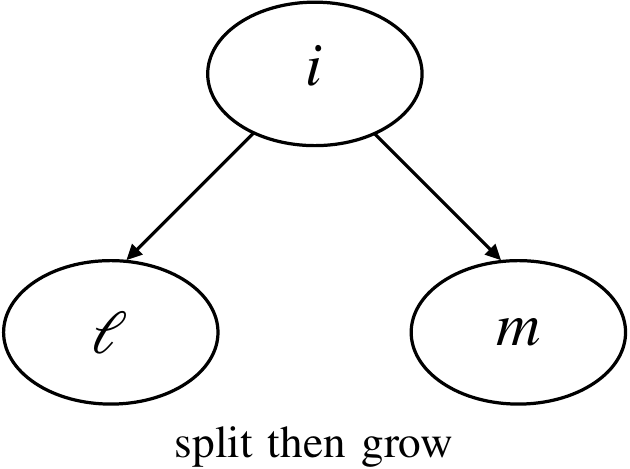}\hspace{0.5cm}
	\includegraphics[width=0.3\textwidth]{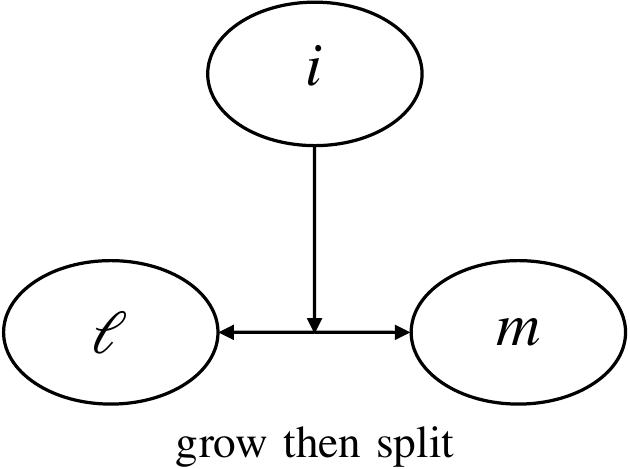}\hspace{0.5cm}
	\includegraphics[width=0.3\textwidth]{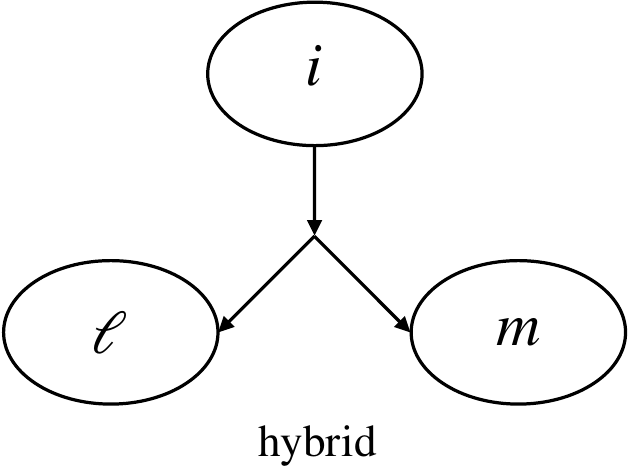}
	\caption{Different prescriptions for the branching time.}
	\label{fig:branching:times}
\end{figure}

In the elementary vertex depicted in \cref{fig:elementary:tree}, as soon as a patch grows larger than twice the Hubble volume, it is divided into two independent patches.
In practice however, the two children patches do not become independent instantaneously, since the decay of gradient interactions at super-Hubble scales is a gradual process.

In stochastic trees, branching is an instantaneous event, but the precise time at which it takes place is not unambiguous.
For instance, instead of splitting a node $i$ and letting the branches leading to the nodes $\ell$ and $m$ grow independently, one could let the node $i$ grow to twice its volume and then split it.
More generally, one could consider a hybrid process where the node $i$ grows up to a time $\alpha\Delta N$, then splits, and the two child branches are evolved independently for $(1-\alpha)\Delta N$, where $0\leq \alpha\leq 1$.
These three possibilities are depicted in \cref{fig:branching:times}.
The ``split-then-grow'' procedure corresponds to $\alpha=0$, while the ``grow-then-split'' procedure corresponds to $\alpha=1$.\footnote{Note that \cref{eq:volume:leaf} was written with the ``split-then-grow'' convention for explicitness. \label{footnote:split_then_grow}}
\begin{figure}[t]
	\centering
	\includegraphics[width=0.99\textwidth]{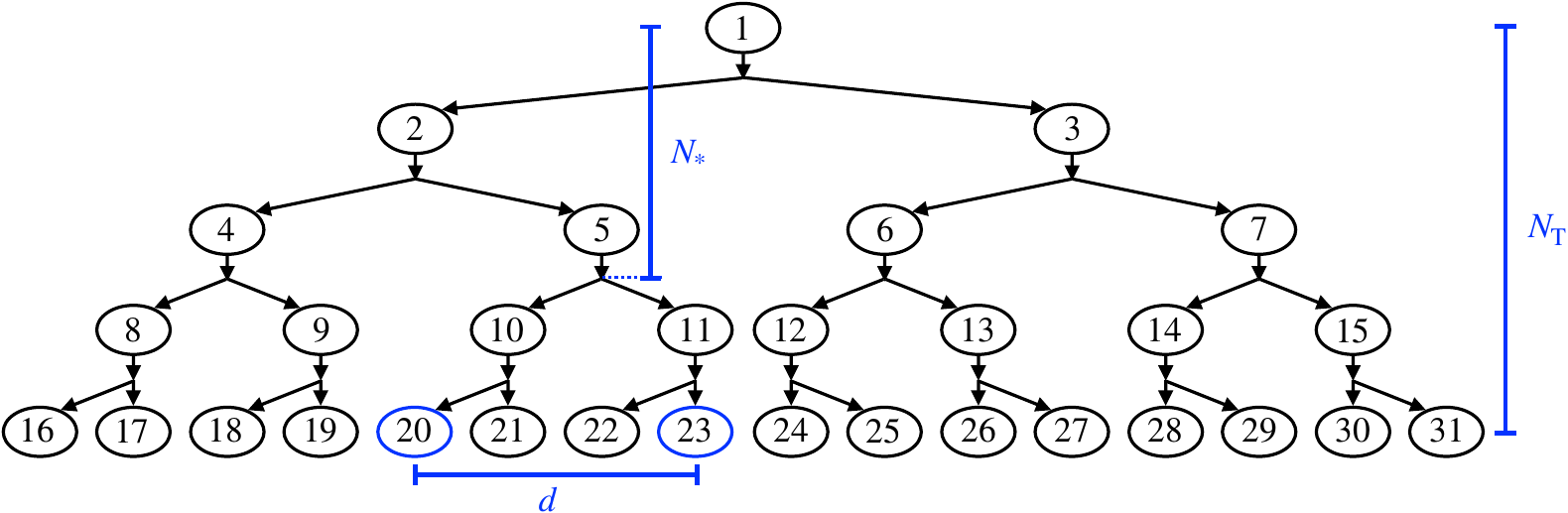}
	\caption{Example tree considered in \cref{sec:discretization:artefact}.}
	\label{fig:example:artefact}
\end{figure}

In order to show explicitly that large-scale properties do not depend on the detailed choice of the $\alpha$ parameter, let us consider the tree depicted in \cref{fig:example:artefact}.
Note that a subscript $\mathrm{T}$ signals quantities associated with the whole tree.
After a time $N_{\mathrm{T}}= q_{\mathrm{T}} \Delta N$ has elapsed from the parent node $1$, consider two nodes $i$ and $j$ separated by a tree distance $d=\abs{j-i}$.
By tree distance, we mean a one-dimensional distance on the final hypersurface in the representation of the binary tree in \cref{fig:example:artefact}, where two sibling nodes are separated by $d=1$.
The maximal such distance one can consider is $D=2^{q_{\mathrm{T}}}-1$.

The paths leading to $i$ and $j$ coincide until node $n_*$, and then become independent.
In the example depicted in \cref{fig:example:artefact}, for $i=20$ and $j=23$, one has $q_{\mathrm{T}}=4$, $d=3$ and the last common ancestor has label $n_*=5$.
First, consider the case where, from $n_*$, the path leading to $i$ always proceeds leftwards, while the path leading to $j$ always proceeds rightwards (we will show how to go beyond this assumption in \cref{sec:branching:surface}).
In this case, the last common ancestor has label $n_*=i/(d+1)$.\footnote{
    More generally, it is convenient to use the binary representation of the nodes to identify the last common ancestor.
    For example, $i = 20 = 10100_{2}$ and $j = 23 = 10111_2$, their last common ancestor is the longest sequence of identical bits in the two numbers, so $101_2 = 5$.
}
The duration of the common path between nodes $i$ and $j$ is thus
\begin{equation}
	N_*=N_{\mathrm{T}}-\Delta N_*
	\quad\text{where}\quad
	\Delta N_* = \left[\frac{\ln(d+1)}{\ln(2)}-\alpha\right]\Delta N\, .
	\label{eq:DeltaNstar:d:alpha}
\end{equation}

Let us now denote the solution of the Fokker-Planck equation associated to \cref{eq:Langevin} by $P(\phi\vert\phi_\mathrm{in},N)$.
This is the probability that, starting from $\phi_\mathrm{in}$, the test field takes value $\phi$ at time $N$.
The joint probability for the field values in the nodes $i$ and $j$ is given by
\begin{equation}
\label{eq:two:pt:distrib:marg:phistar}
	P(\phi_i,\phi_j) = \int \dd{\phi_*}  P(\phi_* \vert \phi_1,N_*) P(\phi_i \vert \phi_*,\Delta N_*) P(\phi_j \vert \phi_*,\Delta N_*)
\end{equation}
where $\phi_*$ denotes the field value at the branching point.
When $1\ll d\ll D$, changing $\alpha$ amounts to changing $N_*$ and $\Delta N_*$ by a small amount and this has little effect on the correlations between $\phi_i$ and $\phi_j$.
This can be more easily seen by rewriting the above result in terms of the three-dimensional physical distance $d_{\mathrm{P}}$ between the nodes $i$ and $j$.
The set of nodes $i$, $i+1$, $\cdots$, $j-1$, $j$, represents a volume $(d+1) V_\sigma = 4 \pi (d_{\mathrm{P}}/2)^3/3$, which leads to the following relationship between $d$ and $d_{\mathrm{P}}$
\begin{equation}
	\label{eq:d:dP}
	d+1=\left(\frac{d_{\mathrm{P}}\sigma H}{2}\right)^3\, .
\end{equation}
Hence  \cref{eq:DeltaNstar:d:alpha} can be rewritten as
\begin{equation}
	\label{eq:DeltaNstar:dP}
	\Delta N_* = \ln\left( Hd_{\mathrm{P}}\right)+\ln\left(2^{-1-\frac{\alpha}{3}} \sigma\right) .
\end{equation}
As a consequence, since $P(\phi_i,\phi_j)$ depends on $\alpha$ only through $\Delta N_*$,  $\alpha$ can be entirely reabsorbed in the definition of $\sigma$, hence the choice of the branching time is degenerate with the choice of the coarse-graining scale.

For explicitness, let us consider a light test field with $V(\phi)=m^2\phi^2/2$ in a de-Sitter universe with Hubble parameter $H$.
In that case the stochastic problem has Gaussian solution~\cite{Starobinsky:1994bd}
\begin{equation}
	P(\phi\vert\phi_\mathrm{in},N) =  \frac{e^{-\frac{\left[\phi-\bar{\phi}(N,\phi_\mathrm{in})\right]^2}{2 s^2(N)}}}{\sqrt{2\pi s^2(N)}}
\end{equation}
where
\begin{equation}
	\bar{\phi}(N,\phi_\mathrm{in})=\phi_\mathrm{in} e^{-\frac{m^2}{3H^2}N}
	\quad\text{and}\quad
	s^2(N)=\frac{3H^4}{8\pi^2m^2}\left(1-e^{-\frac{2m^2}{3H^2}N}\right)\, .
\end{equation}
From this expression, the two-point distribution function can be obtained with \cref{eq:two:pt:distrib:marg:phistar} and is found to be also Gaussian,
\begin{equation}
	P(\phi_i,\phi_j) = \frac{1}{\sqrt{(2\pi)^2\det\Sigma}}e^{-\frac{1}{2} (\Delta\phi_i , \Delta\phi_j) \cdot \Sigma^{-1} \cdot \begin{pmatrix}  \Delta\phi_i
		\\ \Delta \phi_j\end{pmatrix}}
\end{equation}
where we have introduced $\Delta\phi_i =\phi_i - \bar{\phi}(N_{\mathrm{T}},\phi_1)$ and the covariance matrix $\Sigma$ reads
\begin{align}
	\Sigma_{ii}=  & \left\langle \Delta\phi_i^2\right\rangle=\Sigma_{jj}=\left\langle \phi_j^2\right\rangle = s^2\left(N_{\mathrm{T}}\right)
	\, ,                                                                                                                                          \\
	\Sigma_{ij} = & \left\langle \Delta\phi_i \Delta\phi_j\right\rangle=s^2\left(N_*\right)e^{-\frac{2m^2}{3H^2}\Delta N_*}
	\, .
	\label{eq:Sigma:ij:DeltaNstar}
\end{align}
As mentioned above, changing $\alpha$ yields corrections that are suppressed by $\alpha/\ln(d)$ and which can thus be neglected at large distances (when $N_*\gg H^2/m^2$, these corrections are further suppressed by $m^2/H^2$).

In passing, it is worth noting that the two-point correlation function of a free test field in a de-Sitter background can be computed in full quantum-field theory~\cite{Chernikov:1968zm, Tagirov:1972vv, Bunch:1978yq}.
Once proper renormalization is performed, at late time one finds $\Sigma_{ii} = 3H^4/(8 \pi^2m^2)$ in the coincident limit and $\Sigma_{ij} \propto (H d_{\mathrm{P}})^{-\frac{2m^2}{3H^2}}$ at large distance, which matches the above results.

\subsubsection{Branching surfaces}
\label{sec:branching:surface}

When space is described as a tessellation of Hubble patches, one has to introduce hypersurfaces that separate the patches, but the position and the geometry of those hypersurfaces is somewhat arbitrary.
For instance, two nearby points located on each side of a branching hypersurface may be less distant than the Hubble radius, so treating them as disconnected is not strictly valid.
More fundamentally,
branching hypersurfaces single out specific regions in physical space, which breaks homogeneity of spacetime.
This creates artefacts in our discretization scheme.

This can be readily seen by considering neighbour leaves in the tree represented in \cref{fig:example:artefact}, and their degree of parenthood $\Delta N_* = \Delta q_* \Delta N$.
The leaves $16$ and $17$ are such that $\Delta q_*=1$, \ie they are sibling leaves.
However, leaves $17$ and $18$ are such that $\Delta q_*=2 $, \ie they are cousin leaves, and leaves $19$ and $20$ have $\Delta q_*=3$ while leaves $23$ and $24$ have $\Delta q_*=4 $.
In other words, $\Delta N_*(i,j)$ is not just a function of $\vert i-j\vert$, hence $P(\phi_i,\phi_j)$ is not a function of $\vert i-j\vert$ either, which manifestly breaks space-translation invariance.
The problem is that the topological distance $\Delta q_*$ in the tree is not directly mapped to the geometrical distance $d$ on the final hypersurface.

As we shall now argue, one way to solve this issue is to perform ensemble averages over the leaves (in addition to performing stochastic averages over the tree realizations).
For instance, the two-point correlation function of the field fluctuation at physical distance $d_{\mathrm{P}}$ shall be defined as the ensemble average over all pairs of two leaves distant by $d$ on the final hypersurface (there are $2^{q_{\mathrm{T}}}-d$ such pairs)
\begin{equation}
	\label{eq:Sigma:dP:average}
	\Sigma\left(d_{\mathrm{P}}\right) =
	\frac{1}{2^{q_\mathrm{T}}-d}
	\sum_{i=2^{q_\mathrm{T}}}^{ 2^{q_\mathrm{T}+1}-d-1} \Sigma_{i, i+d}
\end{equation}
where $d$ and $d_{\mathrm{P}}$ are related through \cref{eq:d:dP}.
In the example discussed in \cref{sec:branching:time}, this can be computed as follows.
Amongst the $2^{q_{\mathrm{T}}}-d$ pairs of leaves separated by $d$ at time $N_{\mathrm{T}}=q_{\mathrm{T}}\Delta N$, let $\beta(d,q)$ denote the number of pairs whose $\Delta q_*$ is such that $\Delta q_*=q_{\mathrm{T}}-q$. 
For instance, when $d=1$, there is only one pair of leaves such that $\Delta q_*=q_{\mathrm{T}}$, namely the central pair of leaves.
Likewise, there are two pairs of leaves such that  $\Delta q_*=q_{\mathrm{T}}-1$, so on and so forth, hence $\beta(1,q)=2^q$.
One can show that in general, the number of pairs reads
\begin{equation}
	\beta(d,q) =
	\begin{cases}
		2^q d\quad\text{if}\quad d\leq 2^{q_{\mathrm{T}}-q-1}                                                  \\
		2^{q_{\mathrm{T}}}-2^q d \quad \text{if}\quad  2^{q_{\mathrm{T}}-q-1} \leq d \leq 2^{q_{\mathrm{T}}-q} \\
		0 \quad\text{if}\quad d\geq  2^{q_{\mathrm{T}}-q}
	\end{cases}\, .
	\label{eq:alpha}
\end{equation}
A demonstration involving the binary representation of the nodes is given in \cref{sec:counting}.

This counting function is maximal when $d=2^{q_{\mathrm{T}}-q-1}=2^{\Delta N_*/\Delta N-1}$, which is close to the configuration~\eqref{eq:DeltaNstar:d:alpha} to which the analysis of \cref{sec:branching:time} was restricted.
However, all values of $\Delta q_*$ are encountered in a given tree, and they contribute to the spatially averaged correlation functions.

Since $\Sigma_{i,i+d}$ is only a function of $\Delta N_*$ given by \cref{eq:Sigma:ij:DeltaNstar}, one can organize \cref{eq:Sigma:dP:average} according to
\begin{equation}
	\label{eq:Sigma:dp:interm}
	\Sigma\left(d_{\mathrm{P}}\right) =
	\frac{1}{2^{q_{\mathrm{T}}}-d} \sum_{q=0}^{q_{\mathrm{T}}-1} \beta(d,q) \Sigma\left[\left(q_{\mathrm{T}}-q\right)\Delta N\right]\, .
\end{equation}
Inserting \cref{eq:Sigma:ij:DeltaNstar,eq:alpha} into \cref{eq:Sigma:dp:interm} leads to
\begin{equation}
	\Sigma\left(d_{\mathrm{P}}\right) =
	\frac{3H^4}{8\pi^2m^2}\frac{e^{-a q_{\mathrm{T}}}}{2^{q_{\mathrm{T}}}-d}
	\left[
		2^{q_{\mathrm{T}}} \left(e^{a q_*}-1\right)-\frac{2 \left(e^a-1\right) d \left(2^{q_*} e^{a q_*}-1\right)}{2 e^a-1}\right]
\end{equation}
where we have introduced $a=2 m^2\Delta N/(3H^2)$ and $q_*=\lfloor q_{\mathrm{T}}-\ln(d)/\ln(2) \rfloor$.
At large distances, $1\ll d \ll D$, the above can be expanded in the limit $q_* \gg 1$, and it reduces to
\begin{equation}
	\Sigma\left(d_{\mathrm{P}}\right) 
	\simeq \frac{3 H^4}{8\pi^2 m^2}
	\left[\frac{e^{3a}}{2e^a-1} - e^{3a-a q_*}\right]
	\left(\sigma H d_{\mathrm{P}}\right)^{-\frac{2m^2}{3H^2}}\, ,
\end{equation}
where we have further used \cref{eq:d:dP} to relate $d$ and $d_{\mathrm{P}}$.
In the limit $a\ll 1$, which is required for the slow-roll treatment of the test field to be valid, the term in the square brackets can be written as $1-e^{-\frac{2m^2}{3H^2}N_*}$, where we have used \cref{eq:DeltaNstar:dP} to write $q_* = N_*/\Delta N$.\footnote{One may also absorb the $a$-dependence in the redefinition of $\sigma$.} This coincides with the result obtained in \cref{sec:branching:time}, which is also consistent with quantum-field-theory calculations.

This shows that, even though $\Sigma(d_{\mathrm{P}})$ receives contribution from pairs of leaves with $\ln(\sigma H d_{\mathrm{P}})\leq \Delta N_*(i,j)\leq N_{\mathrm{T}}$, in practice those with $\Delta N_*(i,j) \simeq \ln(\sigma H d_{\mathrm{P}})$ dominate, and the correct result is recovered at large distance.
The discretization scheme could be improved, for instance by randomly drawing the position of the branching hypersurfaces within the patches.
This was proposed in \Refa{Jain:2019gsq} and makes the trees statistically homogeneous indeed.
However, it never produces more ``correctly correlated'' pairs than the standard tree, hence it does not improve the convergence towards the correct result at large distance.

Note that in this work, we have not computed the two-point correlation functions discussed above, but rather consider one-point distributions of the coarse-grained or coarse-shelled curvature perturbation. Since the method outlined in \cref{sec:coarse-shelled} allows one to perform coarse-shelling between two nodes that are direct siblings only, only such pairs of nodes are accounted for in our stochastic averages. Given that these pairs are ``correctly correlated'' in the language developed above, we expect our results for the statistics of PBHs to be even more robust to discretisation artefacts than the two-point correlation function. In any case, in the simple example discussed here (a test free field), Gaussian statistics allows one to express the one-point distribution entirely in terms of the two-point correlation functions, hence the convergence of the latter under proper averaging ensures the convergence of the former.

\section{Conclusion}
\label{sec:conclusion}

In this work, we have implemented the stochastic-$\delta N$ formalism on stochastic trees, conceptualizing the inflationary expansion as a branching process.

In essence, inflationary spacetime can be imagined as a branching tree, starting with a single Hubble patch that after a certain amount of stochastic evolution, described by Langevin equations, doubles in volume, giving rise to two new nodes, each representing a descendant Hubble patch then evolving independently.
This process repeats recursively, with each vertex splitting into further branches until inflation ends. The final nodes, where inflation terminates, form the leaves of the tree, where the curvature perturbation and other cosmological fields are measured.
The statistical properties of these fields are embedded in the tree structure: the distance between two leaves corresponds to the recursive level below which their paths separated, \ie to the depth of their latest common ancestor.

This intricate structure can be numerically realized with \texttt{FOREST}~\cite{auclair_2025_15235932}, a cutting-edge, parallel code that generates and scans vast populations of stochastic trees to deliver relevant statistics.
We showed how \texttt{FOREST} returns the first-passage time, in terms of \efolds, through the end-of-inflation hypersurface, averaged over the volume emerging from the initial node, and how the coarse-grained curvature perturbation at the Hubble scale at the end of inflation, or at any arbitrary scale, can be reconstructed.

Stochastic trees tessellate spacetime into Hubble patches that evolve and emerge from branches, making them discrete objects in both time and space.
We presented how the information contained in stochastic trees can be translated into real-space maps of the end-of-inflation hypersurface, unfolding a comoving volume by progressively dividing it along orthogonal planes with alternating orientations at each volume-doubling time.
Despite arbitrariness in  the convention employed to discretize space and in the choice of a branching time, translation invariance and consistency with quantum field theory in curved spacetime are recovered at large distances when large populations of trees are considered.

Stochastic trees are also ideal tools for harvesting primordial black holes.
In addition to the curvature perturbation, we have shown how they allow for the reconstruction of the coarse-shelled curvature perturbation, namely, the curvature perturbation averaged between two concentric spheres.
This serves as a proxy for the compaction function, a cosmological field that is better suited for studying the formation of primordial black holes due to its local nature.
We have shown how it can be viewed as an ``unbalance index'': primordial black holes tend to form at the most asymmetric nodes of the tree.
Moreover, the cloud-in-cloud problem, in which a hierarchy of overdense regions could lead to PBH formation inside larger PBHs, is directly implemented within \texttt{FOREST}, which first scans the smallest scales to check for PBH formation, propagating from the leaves to the root across the tree and discarding black holes when a larger one is found in the same path.

We applied \texttt{FOREST} to a simple toy model, namely the ``quantum well'', which had previously been studied using other methods. 
We reconstructed the probability distribution for the final volume and for the volume-averaged expansion.
Both exhibit a transient power-law behaviour turning to exponential in the far tail.
Tails become heavier as $\mu$ -- the order parameter measuring the amount of quantum diffusion --  increases, up to a critical value which sets the onset of eternal inflation.
Close to the critical point, the volume distribution features a power-law tail and the mean volume diverges.

When scanning the set of leaves, we also observed that exponential tails appear in their first-passage-time distribution. Remarkably, this distribution is perfectly matched by the volume-weighted first-passage-time distribution drawn from independent Langevin processes, which can be computed analytically and which is central to various formulations and approximations of stochastic inflation in terms of first-passage-time analysis.

We have reconstructed the mass distribution of PBHs in the quantum well.
We found that the typical mass of PBHs increases with $\mu$ and that the distribution itself exhibits mild power laws terminated by exponential tails and that become broader as $\mu$ increases. Near the critical point, nearly all the universe is contained into PBHs.

Finally, we discussed how these results compare with those obtained using the previously developed large-volume approximation, which replaces the final volume with its stochastic average.
We found that, for the total PBH mass fraction, the simple Press-Schechter estimate based on the curvature perturbation compares well with the full numerical results (in spite of neglecting cloud in cloud, using the large-volume approximation and relying on the curvature perturbation instead of the density contrast). This shows that the usual estimates of the PBH abundance performed in the stochastic-inflation literature provide reasonable approximations.

However, we also uncovered that the details of the mass distribution cannot be correctly captured by these approximations. The main reason is that the large-volume approximation connects physical scales at the end of inflation to field values during inflation via the mean volume. Yet, PBHs form at the root of subtrees that generate an anomalously large volume, very far from the mean volume indeed. As a consequence, they form at scales and masses that are much larger than what the large-volume approximation predicts.
Since PBHs are extreme phenomena living in the ``tail'' of distribution functions, it should probably not come as a surprise that they cannot be described by a central limit theorem.
Note that this does not preclude the large-volume approximation to perform well for bulk statistics such as the two- or three-point correlation functions, and we defer its comparison with stochastic trees in models producing large volumes to future investigations.

\bigskip
Before concluding, let us highlight several points that undoubtedly deserve attention and offer potential for future development.

By simulating stochastic trees, \texttt{FOREST} reconstructs statistics through direct sampling, enabling, for instance, the investigation of the first-passage-time distribution down to values of probability as small as $10^{-10}$, improving on the numerical efficiency of direct Langevin simulations and lattice codes. 
To investigate the tails of the distributions even more efficiently, methods based on importance sampling can be employed~\cite{Jackson:2022unc, Jackson:2024aoo}, and we leave this possibility for future versions of \texttt{FOREST}.

In this work, we focused on the simplest quantum-well toy model, which, despite its unrealistic simplicity, served as an ideal playground for testing and demonstrating the power of \texttt{FOREST}.
We also confined our study to slow-roll inflation.
However, achieving enhanced perturbations at small scales that could lead to primordial black hole formation typically requires departing from the slow-roll regime and incorporating non-attractor phases, such as ultra-slow roll. We thus plan to extend 
\texttt{FOREST} to higher-dimensional phase spaces that include the field's momentum, and to investigate more realistic models in the future.

In this study, we employed \texttt{FOREST} to reconstruct one-point statistics and make predictions for the abundance and mass distribution of PBHs.
However, its potential extends beyond this, being capable to tackle multiple-point statistics.
The information embedded in the topological structure of stochastic trees can map real-space correlations of cosmological fields at the end of inflation.
This would allow for a non-perturbative reconstruction of the power spectrum and higher-order correlations, and offer a way to explore the spatial distribution of PBHs at formation.
In this way we could check whether PBHs created from the collapse of large non-Gaussian fluctuations are naturally born clustered, as suggested by recent analytical investigations~\cite{Animali:2024jiz}, and if they display an exclusion effect on small-scales~\cite{Auclair:2024jwj}.
This is the subject of a forthcoming publication.

Finally, $N$-body simulations heavily rely on initial conditions drawn from inflationary models.
In this context, stochastic trees emerge as an ideal tool to provide non-perturbative initial conditions in the form of ready-to-use maps. This will help us understanding how PBHs~\cite{Siles:2024yym}, and more generally heavy-tailed statistics~\cite{Coulton:2024vot}, affect structure formation. This offers a fascinating perspective for bridging the physics of the early and late universe through comprehensive, full-scale simulations.

\begin{acknowledgments}
    The authors thank Marie-Paule Faber, Christophe Ringeval and Juan Garc\'ia-Bellido for insightful discussions.
	Pierre Auclair is a Postdoctoral Researcher of the Fonds de la Recherche Scientifique – FNRS.
	Baptiste Blachier is a Research Fellow of the Fonds de la Recherche Scientifique – FNRS.
	Chiara Animali is supported by the ESA Belgian Federal PRODEX
	Grant $\mathrm{N^{\circ}} 4000143201$.
	Computational resources have been provided by the CURL's development cluster, the supercomputing facilities of the Universit\'e catholique de Louvain (CISM/UCL) and the Consortium des \'Equipements de Calcul Intensif en F\'ed\'eration Wallonie Bruxelles (C\'ECI) funded by the Fond de la Recherche Scientifique de Belgique (F.R.S.-FNRS) under convention 2.5020.11 and by the Walloon Region.
\end{acknowledgments}

\appendix

\section{Scaling tests}
\label{sec:scaling}

\begin{table}[t]
    \centering
    \begin{tabular}{rrr | rrr}
        \multicolumn{3}{c}{One node} & \multicolumn{3}{c}{Two nodes (MPI)} \\
        \hline
        \hline
        \# of threads & Wall-time & Speed-up & \# of threads & Wall-time & Speed-up  \\
        \hline
        $1$ & 02:23:30 & $-$ & $-$ & $-$ & $-$ \\
        $2$ & 01:29:46 & $1.6$ & $2 \times 1$ & 01:13:56 & $1.9$ \\
        $4$ & 00:44:26 & $3.2$ & $2 \times 2$ & 00:44:30 & $3.2$ \\
        $8$ & 00:22:15 & $6.5$ & $2 \times 4$ & 00:22:12 & $6.5$ \\
        $16$ & 00:11:14 & $13$ & $2 \times 8$ & 00:11:10 & $13$ \\
        $32$ & 00:05:45 & $25$ & $2 \times 16$ & 00:05:38 & $25$\\
        $64$ & 00:03:00 & $48$ & $2 \times 32$ & 00:02:52 & $50$ \\
        $128$ & 00:01:48 & $80$ & $2 \times 64$ & 00:01:30 & $95$\\
        $256$ & 00:00:56 & $153$ & $2 \times 128$ & 00:00:54 & $159$ \\
        $-$ & $-$ & $-$ & $2 \times 256$ & 00:00:28 & $308$
    \end{tabular}
    \caption{
        Strong-scaling test of \texttt{FOREST} in the flat-well model with $\mu=0.8$ and $10^8$ trees.
        We test the code on two EPYC CPUs communicating through MPI.
        On each computing node, we use OpenMP to distribute the work on the $128$ cores / $256$ threads.
        Times are reported in the format HH:MM:SS, which is hours, minutes and seconds.
    }
    \label{tab:strong-scaling}
\end{table}

\begin{table}[t]
    \centering
    \begin{tabular}{rrrr}
        \# of threads & \# of trees & Wall-time & Efficiency \\
        \hline
        \hline
        $1$ & $10^6$ & 01:27 & $-$ \\
        $2$ & $2 \times 10^6$ & 01:47 & $81\%$ \\
        $4$ & $4 \times 10^6$ & 01:47 & $81\%$ \\
        $8$ & $8 \times 10^6$ & 01:47 & $81\%$ \\
        $16$ & $16 \times 10^6$ & 01:49 & $79\%$ \\
        $32$ & $32 \times 10^6$ & 01:51 & $79\%$ \\
        $64$ & $64 \times 10^6$ & 01:55 & $75\%$ \\
        $128$ & $128 \times 10^6$ & 02:17 & $63\%$ \\
        $256$ & $256 \times 10^6$ & 02:25 & $60\%$ \\
    \end{tabular}
    \caption{
        Weak-scaling test of \texttt{FOREST} in the flat-well model with $\mu=0.8$ on a single EPYC CPU containing $128$ cores and $256$ threads.
        Times are reported in the format MM:SS, which is minutes and seconds.
    }
    \label{tab:weak-scaling}
\end{table}

The main task of \texttt{FOREST} is to simulate a large population of independent stochastic trees to sample the statistics of the end-of-inflation hypersurface.
Our strategy to parallelize the code is straightforward: we first distribute our samples evenly across the different nodes on the cluster using MPI.
On each node, we use OpenMP to distribute the workload on the different threads.
This hybrid approach using MPI+OpenMP is becoming a standard, and combines the advantages of MPI -- communication between an array of distinct computers -- and of OpenMP -- better optimization on machines with shared memory and advanced load balancing.
Each thread contains its own copy of the variables and objects in the code to avoid a memory access bottleneck.
Threads run independently without any communication and results are combined at the end of the simulation using a sum reduction.

We report \emph{strong} scaling results in \cref{tab:strong-scaling}, in which we increase the number of threads for a given task.
Strong scaling is an indicator of whether the wall-time of a given simulation can be reduced by increasing the number of parallel processes.
In the ideal limit, the speed-up factor should equal the number of processes.
As a benchmark, we use the flat-well potential with $\mu=0.8$ and $10^8$ samples.
We find that, apart from a small overhead due to the usage of OpenMP, our code scales consistently from $2$ to $64$ threads per node.
In contrast, combining different nodes with MPI does not produce any overhead.
The reason for such a difference in behaviour is that on a single node, OpenMP has to distribute the shared memory to the different threads.
Depending on the node topology and on the memory allocation strategy, the various processes on the same node can see their memory access times increase significantly.
At $128$ threads per node and above, we still gain a substantial speed-up but at a slower rate.
This behaviour is expected as we saturate the core count of our nodes and may encounter hardware limitations such as power and temperature.

Then, we report \emph{weak} scaling results in \cref{tab:weak-scaling}, in which we increase both the number of threads and the size of the population at the same rate.
Weak scaling is an indicator of whether larger populations can be simulated efficiently by increasing the number of parallel processes.
In the ideal limit, all the tests should have the same wall-time, and we define the efficiency as the ratio between the ideal and the real wall-time.
As a benchmark, we again use the flat-well potential with $\mu=0.8$ and increasing the number of samples from $10^6$.
The conclusions are similar to our strong-scaling tests.
We experience a small drop in efficiency due to OpenMP and a stable efficiency from $2$ to $64$ cores.
Above this threshold, we also note a drop in efficiency.

\section{Convergence tests}
\label{sec:convergence}

\begin{figure}[t]
    \centering
    \begin{subfigure}[t]{.49\textwidth}
        \includegraphics[width=\textwidth]{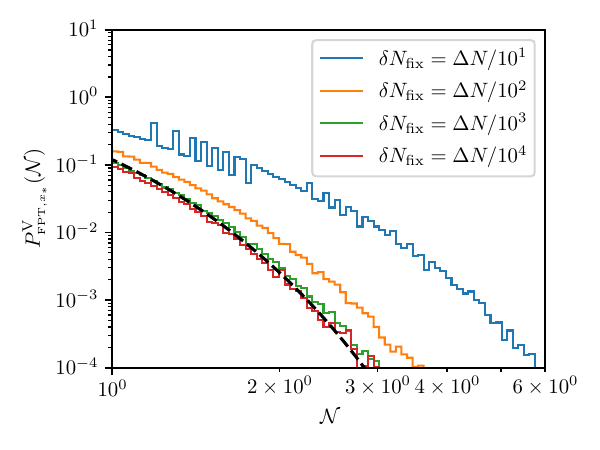}
        \caption{$\mu=0.6$, $10^7$ trees}
    \end{subfigure}
    \begin{subfigure}[t]{.49\textwidth}
        \includegraphics[width=\textwidth]{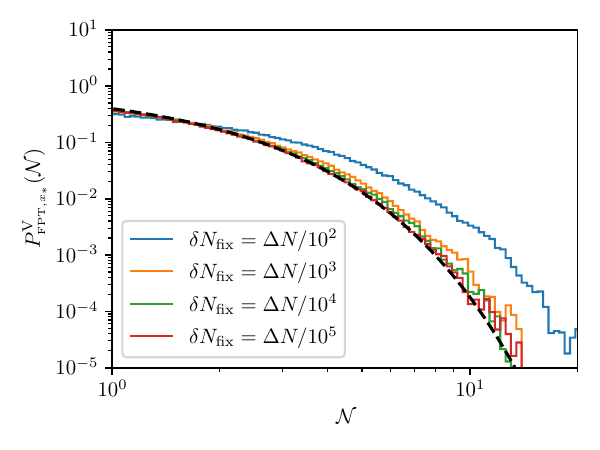}
        \caption{$\mu=0.8$, $10^5$ trees}
        \label{fig:convergence-fixed-0.6}
    \end{subfigure}
    \caption{
        Convergence test for different values of a fixed step $\delta N_\mathrm{fix}$ in the flat quantum well.
        Black dashed line corresponds to the analytical formula of \cref{eq:FPTVolWeight}.
    }
    \label{fig:convergence-fixed}
\end{figure}

\begin{figure}[t]
    \centering
    \begin{subfigure}[t]{.49\textwidth}
        \includegraphics[width=\textwidth]{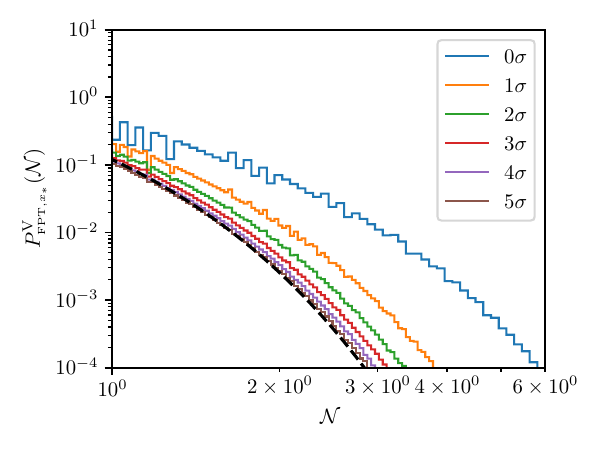}
        \caption{$\mu=0.6$, $10^7$ trees}
    \end{subfigure}
    \begin{subfigure}[t]{.49\textwidth}
        \includegraphics[width=\textwidth]{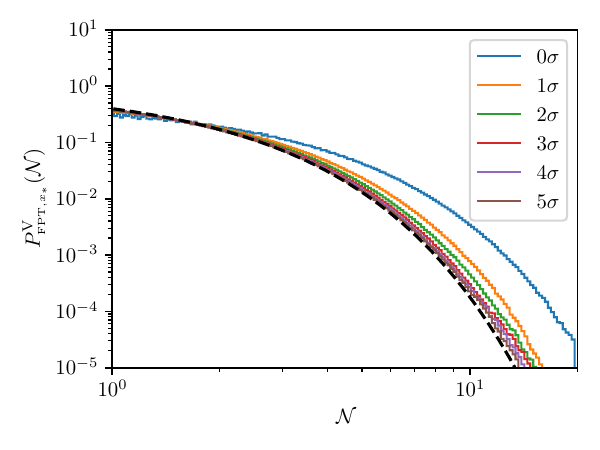}
        \caption{$\mu=0.8$, $10^7$ trees}
        \label{fig:convergence-fixed-0.8}
    \end{subfigure}
    \caption{
        Convergence test for different values of a varying step $\delta N$ following \cref{eq:sigma-step} in the flat quantum well.
        For illustrative purposes, we add the constraint of \cref{eq:sigma-step} to the worst corresponding scenario of \cref{fig:convergence-fixed}.
        That is $\delta N_\mathrm{fix} = \Delta N /10$ for the left panel and $\delta N_\mathrm{fix} = \Delta N / 100$ on the right panel.
        Black dashed line corresponds to the analytical formula of \cref{eq:FPTVolWeight}.
    }
    \label{fig:convergence-sigma}
\end{figure}

At the core of \texttt{FOREST}, we solve the Stochastic Differential Equation (SDE) \eqref{eq:Langevin} using the Euler-Maruyama method.
As briefly mentioned in \cref{sec:tree:structure}, we decided to use a varying step $\delta N$ such that there is at most a $\kappa \sigma$ probability to cross the end-of-inflation hypersurface,
\begin{equation}
    \delta N =
    \min \left\{
    \delta N_\mathrm{fix},
    \frac{3 \left[2\pi \Mp (\phi - \phi_\mathrm{end})\right]^2}{\kappa  V(\phi)}
    \right\}\, ,
    \label{eq:sigma-step}
\end{equation}
in which $\delta N_\mathrm{fix}$ is a fixed value.
The purpose of this appendix is to show, with two quantitative examples, the limitations of using a fixed time step $\delta N = \delta N_\mathrm{fix}$ and why taking $\kappa=5$ is optimal.
As benchmarks, we use the flat-well model, with $\mu=0.6, 0.8$.
Despite its simplicity, the flat-well model can be considered as the limiting scenario in which quantum diffusion dominates completely the dynamics of the system, hence a perfect candidate to study the convergence of our SDE solver.
We focus on the distribution of the volume-averaged elapsed number of \efolds, $\Pfpt{\mathbf{\Phi}_0}^{\mathrm{V}}(\mathcal{N})$, for two reasons.
First, because a value of $\delta N$ that is too large has a very visible impact on this distribution.
Second, because we have an exact analytical formula~\eqref{eq:FPTVolWeight} at our disposal to compare our simulations with.

In \cref{fig:convergence-fixed}, we show the effect of a fixed step $\delta N_\mathrm{fix}$ given as a fraction of the splitting time $\Delta N = \ln(2) / 3$.
If the value of $\delta N$ is too large, then the solver is blind to trajectories that may have exited the quantum well and re-entered during $\delta N$.
Therefore, this tends to over-estimate the first-passage time, as can be seen in \cref{fig:convergence-fixed}.

In \cref{fig:convergence-sigma}, we show how our prescription of \cref{eq:sigma-step} converges when added to the worst-case scenario of \cref{fig:convergence-fixed}.
In both our benchmarks, we find that the prescription~\eqref{eq:sigma-step} substantially improves the convergence of the distribution even for small values of $\kappa$.
This convergence is robust and less sensitive to the specific value of $\mu$ one considers.

Notably, adding our condition~\eqref{eq:sigma-step} to an existing fixed time step does not increase the computation time by more than a factor of $2$ or $3$.
This should be compared with the cost of dividing the size of the time step by several orders of magnitude, and hence multiplying the wall-time by the same amount.
Qualitatively for $\mu=0.8$, it takes about the same time to obtain the best curve in \cref{fig:convergence-fixed-0.6} with $10^5$ trees and the best curve in \cref{fig:convergence-fixed-0.8} with $10^7$ trees, a performance boost of a factor $100$.
The reason for such a performance gain is that reducing the time step increases the resolution of the SDE solver across the whole quantum well.
To the contrary, \cref{eq:sigma-step} increases the resolution close to the end-of-inflation hypersurface only.
Eventually, what we are interested in is the duration of inflation in the different patches and not the full history of the Langevin trajectories.

In production, we impose $\delta N_\mathrm{fix} = 10^{-3}$, a value more stringent than the two benchmarks presented in this appendix.
The remarkable agreement between our large numerical simulations and the corresponding analytical formula can be seen in \cref{Fig:PZetaCompFlat}.

\section{Counting function}
\label{sec:counting}

We consider the balanced binary trees studied in \cref{sec:discretization:artefact}. The goal of this appendix is to compute the counting function $\beta(d,q)$, which corresponds to the number of pairs of leaves $i$ and $j$ such that $j=i+d$ and such that the last common ancestor to the two leaves lies at $N=q\Delta N$ from the root.

\begin{figure}
    \centering
    \includegraphics[width=.2\textwidth]{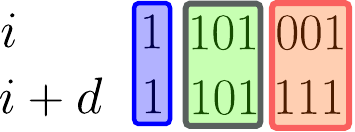}
    \caption{Demonstration for \cref{eq:alpha} using of the binary representation of nodes $i$ and $j = i+d$. In this example, $i=1101001_2 = 105$ and $j= 1101111_2=111$ are distant by $d=6$. One has $q = 3$ (three green bits) and $q_\mathrm{T} - q = 3$ (three red bits).
    }
    \label{fig:pair-proof}
\end{figure}

Following the example displayed in \cref{fig:pair-proof}, the binary representations of $i$ and $j$ are given by strings of bits that contain three segments. 
The first segment, displayed in blue in \cref{fig:pair-proof}, is a single bit, $1$, which represents the root node. 
It is always present and equal to one. 

The second segment, displayed in green, represents the common part of the paths leading to $i$ and $j$, prior to their last common ancestor. 
Its contains $q$ bits that are identical for $i$ and $j$, there are thus $2^q$ possible green segments.
In the tree representation, these correspond to the paths leading to the $2^q$ nodes located $q$ levels below the root node.

The third segment, displayed in red, represents the part of the two paths that lies below the last common ancestor. 
It contains $q_\mathrm{T}-q$ bits in the representation of $i$, which fully determine the bits in $j$ since $j=i+d$. 
Therefore, there are at most $2^{q_\mathrm{T}-q}$ possibilities for the red segment.
However, two conditions need to be fulfilled by the red segment.

First, one must ensure that no green bit is being flipped when summing $i$ and $d$, since by definition the green segments have to remain identical for $i$ and $j$.
This limits the last $q_\mathrm{T}-q$ bits of $i$ to be strictly smaller than $2^{q_\mathrm{T}-q} - d$, since larger values would induce a bit flip.
Therefore, there are $2^q \times (2^{q_\mathrm{T}-q} - d) = 2^{q_\mathrm{T}} - d 2^q$ pairs sharing \emph{at least} the same first $1 + q$ bits, that is having \emph{a} common ancestor at $N=q\Delta N$.
In terms of the counting function, 
\begin{equation}
	\sum_{q' \geq q} \beta(d, q') = 2^{q_\mathrm{T}} - d 2^q \, .
	\label{eq:beta-sum}
\end{equation}
Second, we have imposed that $i$ and $j$ share the same path \emph{at least} until a node at level $q$, but this does not guarantee this node is the \emph{last} common ancestor. 
One must also ensure that $i$ and $j$ follow different branches below this node.
In terms of the binary decomposition of $i$ and $j$, we need to impose that the first bit of the red segment differs between $i$ and $j$.
Three cases need to be distinguished:
\begin{itemize}
    \item If $d \geq 2^{q_{\mathrm{T}}-q}$, then the binary representation of $d$ contains more bits than the red segment, hence adding $d$ to $i$ necessarily flips bits in the green segment. 
	Since this is impossible, $\beta(d, q)=0$ in that case.
    Consistently, \cref{eq:beta-sum} does not give a positive number of pairs in this case.
    \item If $d < 2^{q_{\mathrm{T}}-q-1}$, then we obtain the number of pairs having their last common ancestor at level $q$ using \cref{eq:beta-sum},
    \begin{equation}
		\beta(d, q) 
		= \sum_{q' \geq q} \beta(d, q') - \sum_{q' \geq q + 1} \beta(d, q')
		= 2^q d \, .
	\end{equation}
    \item Finally, if $2^{q_{\mathrm{T}}-q-1} \leq d < 2^{q_{\mathrm{T}}-q}$, then it is not possible to have $q+1$ identical bits in the green segment, hence $\beta(d, q+1) = 0$ and
    \begin{equation}
		\beta(d, q) = \sum_{q' \geq q} \beta(d, q') = 2^{q_\mathrm{T}} - d 2^q \, .
	\end{equation}
\end{itemize}
    
To summarize, we have found that 
    \begin{equation}
	\beta(d,q) =
	\begin{cases}
		2^q d\quad\text{if}\quad d < 2^{q_{\mathrm{T}}-q-1}                                                  \\
		2^{q_{\mathrm{T}}}-2^q d \quad \text{if}\quad  2^{q_{\mathrm{T}}-q-1} \leq d < 2^{q_{\mathrm{T}}-q} \\
		0 \quad\text{if}\quad d\geq  2^{q_{\mathrm{T}}-q}
	\end{cases}\, .
\end{equation}
This is identical to \cref{eq:alpha}, where strict inequalities on $d$ have been changed to weak inequalities after checking that the counting function is continuous at the two pivotal points $d=2^{q_{\mathrm{T}}-q-1} $ and $d=2^{q_{\mathrm{T}}-q}$. 

As a consistency check, one can compute the total number of pairs $i$ and $j=i+d$, regardless of their topological distance $q$. This number is given by
\begin{align}
\sum_{q=0}^{q_{\mathrm{T}}-1} \beta(d,q)
= & \sum_{q=0}^{q_{\mathrm{T}}-2- \lfloor \frac{\ln(d)}{\ln(2)} \rfloor } 2^q d
+\sum_{q=q_{\mathrm{T}}-1- \lfloor \frac{\ln(d)}{\ln(2)} \rfloor}^{q_{\mathrm{T}}-1- \lfloor \frac{\ln(d)}{\ln(2)} \rfloor}\left(2^{q_{\mathrm{T}}}-2^q d\right)
+\sum_{q=q_{\mathrm{T}}- \lfloor \frac{\ln(d)}{\ln(2)} \rfloor}^{q_{\mathrm{T}}-1} 0 \\
= & 2^{q_{\mathrm{T}}}-d\, ,
\end{align} 
where the first sum is over terms of a geometric series and can thus be readily evaluated. One recovers the  total $2^{q_{\mathrm{T}}}-d$ pairs of leaves separated by $d$ at time $N_{\mathrm{T}}$.

\bibliographystyle{JHEP}
\bibliography{TreenPBH}

\end{document}